# Emergence of cellular nematic order is a conserved feature of gastrulation in animal embryos

Xin Li[1*], Robert J. Huebner[2*], Margot L.K. Williams[3,4], Jessica Sawyer[5,6], Mark Peifer[6], John B. Wallingford[2], and D. Thirumalai[1,7]

1. Department of Chemistry, University of Texas at Austin, Austin, TX 78712, USA

2. Department of Molecular Bioscience, University of Texas at Austin, Austin, TX 78712, USA

3. Center for Precision Environmental Health & Department of Molecular and Cellular Biology, Baylor College of Medicine, Houston, TX 77030, USA

4. Department of Developmental Biology, Washington University School of Medicine, St. Louis, MO 63110, USA

5. Department of Pharmacology and Cancer Biology, Duke University, Durham, NC 27710, USA

6. Department of Biology, University of North Carolina at Chapel Hill, Chapel Hill, NC 27599-3280, USA

7. Department of Physics, University of Texas at Austin, Austin, TX 78712, USA

* e-mails: xinlee0@gmail.com, roberth@utexas.edu

# Abstract

Cells undergo dramatic morphological changes during embryogenesis, yet how these changes affect the formation of ordered tissues remains elusive. Here, we show that a phase transition leading to the formation of a nematic liquid crystal state during gastrulation in the development of embryos of fish, frogs, and fruit flies occurs by a common mechanism despite substantial differences between these evolutionarily distant animals. Importantly, nematic order forms early before any discernible changes in the shapes of cells. All three species exhibit similar propagation of the nematic phase, reminiscent of nucleation and growth mechanisms. The spatial correlations in the nematic phase in the notochord region are long-ranged and follow a similar power-law decay ($y{\sim}x^{-\alpha}$) with $\alpha$ less than unity, indicating a common underlying physical mechanism. To explain the common physical mechanism, we created a theoretical model that not only explains the experimental observations but also predicts that the nematic phase should be disrupted upon loss of planar cell polarity (frog), cell adhesion (frog), and notochord boundary formation (zebrafish). Gene knockdown or mutational studies confirm the theoretical predictions. The combination of experiments and theory provides a unified framework for understanding the potentially universal features of metazoan embryogenesis, in the process shedding light on the advent of ordered structures during animal development.

# Introduction

In animal embryos, the evolutionarily conserved process of convergent extension (CE) narrows and elongates the body during gastrulation, transforming the largely spherical early embryo into the elongated animal body plan with a recognizable head-to-tail axis[1]. CE is driven by directed movement of cells in one axis (convergence), resulting in extension of the perpendicular axis[2,3]. A wealth of descriptive and experimental embryology has characterized the cell biology, molecular controls, and mechanics that underlie CE, across species[4-8]. However, it remains unclear how cellular scale behaviors translate to larger-scale physical transitions. In addition, the common (possibly universal) aspects and differences across distantly related species are not known.

Confluent cell populations, such as those found in the early embryo, can exist in a variety of physical states. For example, a high-density tissue consisting of isotropic, non-proliferative, cells will be in a jammed or solid-like state[9]. In contrast, a growing tissue undergoing dramatic morphological changes must be in a more fluid-like or unjammed state[10]. Transitions between such states are influenced by a variety of parameters, including cell density, proliferation, packing order, and, of particular interest here, cell shape[11-14]. Indeed, dramatic changes in cell shape and cell orientation are a defining feature of vertebrate CE, with cells becoming strikingly aligned and elongated in the axis perpendicular to the elongating head-to-tail axis[15-18]. However, the relationship between cell shape, orientation, and the physical state of tissues during CE has not been defined.

Here, we used zebrafish, *Xenopus*, and *Drosophila* as model systems, to consider CE across evolution. In all three species, an ordered phase with orientational order emerges before overt changes in cell shape. Tissues then transition to a nematic liquid crystal phase defined by long-range orientational order. The emergent long-range spatial correlation is characterized by the power-law ($y \sim x^{-\alpha}$), with $\alpha$ being less than unity, during the development of the liquid crystalline phase. Further, we found that each animal spontaneously exhibited propagation of the nematic phase over time and space that could be described by a nucleation-type mechanism. A theoretical model, with only two parameters, suggests a requirement for both local and global cell alignment during nematic phase formation, an idea we tested *in vivo*. The theory explains the effect of mutations and gene knockdowns on the CE. Together, our results provide a physical description of how cells acquire an anisotropic liquid crystal state during early embryogenesis and suggests a potentially universal physical mechanism in the CE of evolutionarily distant animals.

# Results

### Evolution of anisotropic cell shape during zebrafish gastrulation.

Zebrafish elongate their head-to-tail axis, in part, through convergence and extension of the dorsal mesoderm which gives rise to the notochord[4]. Here we began by establishing a quantitative description of known features of convergent extension[17]. At early stages, the presumptive notochord appears as a hexagonally packed confluent tissue[19] (Fig.1a, Supp. Fig.1a-d). Over time, the packing density increases, the mean cell area decreases (Supp. Fig.1e-g and k-m, Supp. Fig.2), and cells become increasingly elongated, particularly in the notochord (dashed lines) (Fig.1b, Supp. Fig.1k-m and Video I). From these expected observations, we began a deeper investigation of the nature of cell anisotropy.

For a more granular view of cell shape changes, we calculated the shape index $SI \equiv P/\sqrt{(4\pi A)}$, where P is the perimeter and A is the area of each cell[14]. This parameter has been used to investigate onset of tissue fluidization because *SI* increases as cells become increasingly anisotropic and the tissue becomes more liquid-like[14,20]. The value of *SI* is unity for a circle and is larger for other shapes. We collected time-lapse movies of zebrafish CE and calculated the *SI* of the cells in the notochord (dashed rectangle) and the surrounding mesoderm (Fig.1a-b). *SI* was plotted as a function of position on the mediolateral axis, perpendicular to the head-to-tail axis, to visualize the spatial evolution of the cell shape (Supp. Fig.1h-j, n-p). We observed a general increase in *SI* across the tissue over time with a particularly strong increase in the notochord region (Supp. Fig.1n-p and Video II). The boxplot in Fig. S1(q) (also Fig. S1(r)) shows that cell shape is significantly different between cells in and outside of the notochord region. There is also an increase in *SI* along the anteroposterior axis over time, but there is no clear pattern as observed along the mediolateral axis (Supp. Fig.3). This result demonstrates the formation of a heterogeneous tissue in which cells with different shapes coexist. Furthermore, the morphology in and out of the notochord region is different. There is a clear orientational order in the notochord, but the cells are essentially disordered outside.

To probe the temporal dynamics of cell shape, we calculated the time dependence of the *SI* for cells in the notochord. Initially, *SI(t)* increases linearly with time, indicating a slow increase in the amplitude of cell shape anisotropy (Fig.1c dark yellow line). However, there is a sharp transition to a steeper slope (an increase of $\sim 50\%$, see also the inset in Fig.1c for the derivative of *SI* over time) around t=120 min (Fig.1c red line), which occurs concurrently with the appearance of the notochord boundary (Supp. Fig.1k-p). There are only minor changes in SI outside of the notochord region, which contrasts with the obvious changes in the *SI* in the notochord region (Supp. Fig.4). These data highlight the sensitivity of the *SI*, as spatial analysis clearly shows a peak in the notochord region and temporal analysis identified boundary formation through a slope change in the rate (Fig. 1c). *SI* is therefore a sensitive parameter for describing the structural transition in tissues during CE *in vivo*. It would be interesting to observe a phase transition from isotropic (*SI*=1) to anisotropic (*SI*>1), if the cells are isotropic at the earliest developmental stages.

**A Nematic phase emerges early in zebrafish gastrulation.**

We next asked how cell alignment arises. We first quantified the orientational order of the cells, which refers to the alignment of anisotropic objects in a particular direction; in this case the arrangement of the long axis of the cell in the tissue (short lines Fig.1d, g at different times). Each line is color-coded by the angle ($|\theta|$, see the color bar and the inset in Fig.1d, Supp. Fig.5a) with respect to the mediolateral axis. The polar histogram of $\theta$ (Supp. Fig.5b) shows a peak around zero degrees, revealing that many cells are aligned along the mediolateral axis even at very early times. This trend becomes even more apparent at later times when almost all the cells in the notochord region orient their long axes along the mediolateral axis (grey dashed rectangle Supp. Fig.5c). The histogram of $\theta$ (Supp. Fig.5d), and the inset (showing the distribution for cells in the dashed rectangle in Supp. Fig.5c) provide additional support for the near perfect orientational order.

The highly oriented cells are reminiscent of the nematic liquid crystal phase. Nematic liquids flow like a liquid (without long-range positional order) while retaining persistent long-range orientational order[21]. Although liquid crystal states have been used to describe biological structures such as DNA, cytoskeletal elements, and even confluent sheets of cultured epithelial cells[22-25] they have not been implicated in tissue transformations during animal development. We quantified the orientational order of cells during zebrafish CE in two dimensions (2D) (Fig.1d, g) by calculating the nematic order parameter[21],

$$S = \langle 2cos^2\theta - 1\rangle, \quad (1)$$

where $\theta$ is the angle shown in Fig. 1(d), and (g). In a perfect nematic liquid, in which all the cells are oriented in the same direction, $S$=1, whereas $S$=0 if the cells are randomly oriented.

Due to the spatial heterogeneity and anisotropic nature of zebrafish tissues, we calculated the value of $S$ along the mediolateral and anteroposterior directions separately (Fig.1e-f, h-i). Strikingly, the presence of a nematic phase is evident at very early times (see the pink line in the center region in Fig.1f) even before there are discernible changes in the shapes of cells. The nematic phase emerges just as the notochord structure is being established. The nematic order in the notochord region is further enhanced at subsequent times, leading to larger $S$ values (approaching unity, see the pink line in the center right region in Fig.1i). The large $S$ value along the anteroposterior axis (Fig.1e) also shows the presence of a nematic phase ($S \approx 0.5$), but its distribution is more homogenous at early times since the notochord spans the entire regime in the anteroposterior direction. Interestingly, there is a gradient in $S$ along the anteroposterior direction at later times (see the blue line in Fig.1h) similar to previous observations of early changes in cell shape[16]. It is worth noting that the value of $S$ is near unity at t = 147 min in the mediolateral region between 100 $\mu m$ and 150 $\mu m$ (Fig. 1i). Even outside this region $S$ has a non-zero value, showing a dramatic development of the nematic order during the CE process.

Our data reveal that cells in the zebrafish notochord transition to a nematic phase during CE (see the well-stacked cells in Fig.1g and $S \approx 1$ in Fig.1i). As an additional test of this finding, we characterized the nematic phase using a two-fold orientational order ($\psi_2$, defined in Eq. (3)), which captures the $180^o$ rotational symmetry (see the Materials and Methods for a definition and the method of calculation). We found that many cells in the notochord region have a large $\psi_2$ (> 0.5, see the dashed rectangle in Supp. Fig.5e). By this measure, the cells in the notochord region are more ordered than the cells on either side in a two-dimensional projection (see Supp. Fig.5f). Together these data firmly establish that the zebrafish notochord displays liquid crystal-like features during CE and prompted us to further explore the generality of this finding.

**The onset of the nematic phase occurs prior to major changes in cellular anisotropy.**

We used the spatiotemporal changes in the alignment of cells to determine the mechanism of formation of the nematic order as CE proceeds. To this end, we first assessed the correlation between the shape index and the associated nematic order parameter using the measured spatiotemporal dependence of *SI* and *S*. At early times, there are no discernible changes in *SI* along the mediolateral axis (Supp. Fig.6a-b). In contrast, a dome-like shape appears in *S* along the same axis, indicating the presence of a nematic phase in the mediolateral region (see Supp. Fig.6d), suggesting that changes in orientational order precede changes in the cell shape. Of interest, microtubules also display polarized growth prior to overt cell shape changes, suggesting a role for the microtubule cytoskeleton in early nematic order formation[26].

The ordered phase in the middle region is stable, with only minor fluctuations in the standard deviation ($\sigma_S$) (see Supp. Fig.6e between the red dashed lines and Supp Fig.6f-h at different times), in contrast to the larger fluctuations in the more disordered region on either side of the putative notochord. Intriguingly, the cells in the nematic phase (enclosed by the two dashed lines) eventually form the zebrafish notochord (Supp. Fig.6c cyan dashed box) even though the ordered notochord structure is not fully established. These observations show that cells change their orientational order early in the CE. Interestingly, at these early times the differences in the cell shapes are modest at best (Supp. Fig.6a-b). In summary, in the zebrafish notochord, *S* and its dispersion, $\sigma_S$ (see more detailed discussion below), are more accurate predictors of the structural and dynamical outcomes during the early stages of CE than the changes in cell anisotropy.

To probe the mechanism of the cell orientation order, we first calculated the time dependence of the nematic order (*S(t)*) for the cells in the notochord region (see Fig.1j). Instead of a gradual increase in the nematic order with time, *S(t)*, grows rapidly (in about 10 minutes it reaches the middle point between the initial value and the final value) and subsequently increases slowly. Because *S(t)* increases substantially before any discernible changes in *SI(t)* (see the minor variation of *SI* values around 1.25 in the inset in Fig.1j with *S(t)* and *SI(t)*), it implies that the transition to the nematic phase in the notochord is unrelated to cell shape changes. We surmise that the tissues rapidly transition to a nematic phase at early times without any correlation between *S(t)* and *SI(t)*. The fluidity of a liquid crystal may facilitate both the CE process and the formation of the notochord (see Video III and Supp. Fig.7 in *SI* for the evolution of cell orientation).

Interestingly, the polydispersity in the cell shape is substantial at early times during which the most rapid changes in *S(t)* are observed (see Fig. S2 for the changes in the area distribution as a function of time). Polydispersity occurs in thermotropic liquid crystals which could undergo phase separation in which long and short chain molecules segregate. Clearly, this is not the case in cells undergoing CE, where no such segregation occurs. In addition, unlike in synthetic liquid crystals[27], the *SI* of cells increases with time together with active cell movement (see the T1 transitions shown in Supp. Fig.8), which may also contribute to the rapid formation of the nematic liquid crystal phase observed in Fig.1j. These results show that an actively flowing nematic phase develops early during zebrafish CE, in which the order parameter *S*, together with its dispersion $\sigma_S$ describe the phase transition in the notochord tissues.

**The nematic phase is established via a nucleation and growth mechanism.**

By analyzing the time dependent changes observed in the images, we investigated how the nematic phase grows and propagates during CE. In Fig.1l, we plot the heatmaps of the order parameter *S* in (I)-(VIII) at successive growth phases. A small "nucleus" with a relatively high *S* value (see the black region enclosed by the white dotted line) forms by t=3 min (possibly earlier) in the center of the field of view in Fig. 1l(I). As time increases, this ordered region grows and expands in both the mediolateral and anteroposterior directions, with a more pronounced expansion in the latter direction, indicative of the start of the extension[2,3]. The boundary of the ordered phase (in black) is initially irregular (see (I)-(V)) and fractal-like. Importantly, fluctuations are observed during the growth of the nematic phase. At later times, a more regular rectangular region forms, representing the zebrafish notochord (see (VI)-(VIII)). The increase in *S(t)* is also accompanied by an increase in the area of the nematic phase, $A_S$. In the notochord region ($S$ > 0.8) we find that $A_S$ increases as -$t^{-0.5}$ (Supp. Fig.9a). A visual representation of the growth of *S* along the anteroposterior axis at different times using a kymograph (see Supp. Fig.9b) shows nucleation and spreading of the nematic order. These findings have the hallmarks of a nucleation and growth mechanism.

**Long range spatial order in the notochord region.**

To characterize the nature of the nematic order in the notochord region, we calculated the spatial correlation $C_S(r)$, describing how the alignment of cells varies with distance. An exponential decay of the function would indicate that alignment weakens rapidly as the distance between the cells increase. However, if $C_S(r)$ decreases slowly, exhibiting a power-law pattern, it suggests there is significant alignment even between cells at large separation, indicating long-range order. The spatial correlation $C_S(r)$ is defined as,

$$C_S(r) = \langle S(r+r_i)\, S(r_i)\rangle / \langle S(r_i)\, S(r_i)\rangle, \qquad (2)$$

where $S(r_i)$ is the nematic order (defined in Eq. (1) without averaging) of a cell at location $r_i$, and the average (<>) is over all cells in the notochord region.

Strikingly, the decrease in the correlation function (Eq. (2)) follows a power law, $C_S(r) \approx r^{-\alpha}$ (Fig.1k) with a small value of $\alpha$, indicating the presence of a long-range spatial correlation in the notochord region. The value of the exponent, $\alpha$, changes in a time-dependent manner, but remains small (see Fig.1k) throughout, which is consistent with the finding that the nematic order in the notochord region emerges early in the CE process and remains persistent (Fig.1f, i).

Taken together, these data reveal that the formation of the zebrafish notochord is accompanied by the growth and propagation of a nematic phase starting from a seed, reminiscent of a nucleation and growth process observed in crystals[28]. Observation of a nematic liquid crystal phase during zebrafish CE prompted us to determine if this liquid crystal state is an evolutionarily conserved feature of CE.

**Evolution of anisotropic cell shape during *Xenopus* CE**.

We next asked if a nematic phase also emerges naturally during *Xenopus* gastrulation. Like zebrafish, *Xenopus* notochord cells undergo CE[3]. But unlike zebrafish, which has an optically accessible notochord, the *Xenopus* notochord is embedded within opaque tissues and is studied ex vivo using explant culture[29-30]. The *Xenopus* notochord is also initially much broader than in zebrafish and takes 8-10 hours to narrow. Here, we visualized the onset of CE. The entire field of view is composed of notochord cells (Fig.2a-b, Supp. Fig.10a-f).

We collected time-lapse movies of *Xenopus* CE and calculated *SI* for the cells. At early times, the presumptive notochord cells (Fig.2a) are much more irregular than in zebrafish (Fig.1a). At later times, the anisotropy of all the cells in the broad notochord increases, as illustrated by both the shape index (*SI*), and the aspect ratio during CE (Supp. Fig.10g-h) and Video IV). The time dependence of *SI* for the cells in the field of view shows that it increases slowly with time (t) and subsequently begins to grow faster (by about 30%, see also the inset in Fig.2c) after t=70 min (orange and red lines in Fig.2c). These data indicate that, as is the case with zebrafish, *SI* captures the structural transition of the tissue during CE.

**The nematic phase and growth mechanism are conserved in *Xenopus*.**

In *Xenopus* there is clear evidence of orientational order at very early times (see Fig.2d-f), though there are differences along the anteroposterior axis. For example, most of the cells at the anterior (top)of Fig.2d are oriented along the mediolateral axis and the orientational order parameter *S* in this region approaches unity (Fig.2e). This is consistent with pioneering observations that CE progresses from anterior to posterior[15]. As one moves posteriorly, there is a region of mixed polarity, followed by a second region of mediolateral orientation, while more posterior cells are more randomly oriented. It follows that the cells in *Xenopus* tissue also start to form a nematic liquid crystal at early times of CE, although the cells have irregular shapes, which appear to be randomly packed (see Supp. Fig.10a). Such an 'active' nematic phase could contribute to the self-assembly of ordered tissue structures, exhibiting both fluid and solid-state like properties simultaneously.

Subsequently, almost all the cells align along the mediolateral axis (Fig.2b)[15], and form a near perfect nematic phase (see Fig.2g-i). Development of the nematic order is also captured by the time dependence of *S* for the cells in the field of view (see Fig.2j). The value of *S* increases rapidly after a short lag time (not observed in Fig.1j for zebrafish). Interestingly, the rapid growth in the nematic order in *Xenopus*, after the lag time, can be described by a function that is similar to that used to analyze the data in zebrafish (compare the solid lines and the function listed in Fig.1j and 2j). It is worth emphasizing that, just as in zebrafish, nematic order develops before significant changes in cell shape occur (see the inset in Fig.2j). The spatial correlation function, $C_S(r)$ (see Fig.2k), also exhibits a power-law behavior at all times, indicative of the presence of a long-range correlation in *S* in the nematic ordered phase. Taken together, the data suggest that the notochord rapidly forms a nematic phase with establishment of a long-range spatial order at early times during CE in both fish and frogs.

To better understand the spatiotemporal evolution the nematic phase, we calculated the heatmaps of *S* at different times (see Fig.2l). Just as in zebrafish tissue (Fig.1l), a small domain with an ordered phase initially forms at the right center of the field of view, and then begins to expand and propagate throughout the tissue in the field of view (see Fig.2l(I)-(VIII)). The greater number of cells in the notochord of *Xenopus* may contribute to the less regular expansion compared to that in zebrafish. Our analyses reveal that a nematic phase emerges locally, then expands by a nucleation-growth process in both zebrafish and *Xenopus* CE.

**Evolution of anisotropic cell shape during *Drosophila* CE.**

Given the finding that a nematic liquid crystal phase emerges spontaneously during gastrulation in two vertebrate organisms, we wondered if a similar behavior is found in a more an evolutionally distant animal. *Drosophila*, zebrafish, and *Xenopus* have been primary organisms used to study convergent extension and qualitative comparisons between them is instructive[3]. While many features of CE are conserved between *Drosophila* and vertebrates[3], there are also significant differences. First, *Drosophila* CE occurs in the epithelial ectoderm, as opposed to the mesenchymal mesoderm in fish/frogs, in a region known as the germ-band[31]. Second, vertebrates and *Drosophila* employ different molecular mechanisms to pattern the converging and extending tissue[7,32-33]. To assess if the physical principles of CE (in particular the emergence of the ordered nematic phase) are evolutionarily conserved despite these substantial differences, we collected time-lapse movies of *Drosophila* CE and used *SI* to analyze cellular anisotropy.

Two snapshots of the *Drosophila* CE are shown in Fig.3a-b. The *Drosophila* tissue extended along the head-to-tail axis (that is the horizontal axis in Fig.3a-b). However, in contrast to zebrafish and *Xenopus*, cells are elongated in the head-to-tail direction (see Fig.3b and Video V). To illustrate the cell shape changes during CE in *Drosophila*, we calculated the time dependence of *SI* for the cells in the dashed rectangle region in Fig.3b. Surprisingly, *SI* initially decreases with time, followed an increase at later times (see the orange and red lines in Fig.3c). In about 10 minutes, the *SI* abruptly reaches a high value (see the inset in Fig.3c), saturating at $SI \approx 1.23$. Thus, *Drosophila* cells do reach an anisotropic steady state during CE. The associated temporal dynamics are substantially different from the other two species. More importantly, the orientation is flipped compared to the zebrafish and *Xenopus*, with the *Drosophila* cells elongating in the same direction as the elongating anterior-posterior axis. These differences prompted us to further explore nematic order during *Drosophila* CE.

**Nematic order is conserved in *Drosophila.***

In the *Drosophila* embryo analyzed in Figure 3 a small nematic domain appears at t=15 minute (the upper left region in Fig.3a; cell orientation in Fig.3d) with $S > 0.5$ (Fig.3e-f). At later times, the nematic order domain expands into the lower anterior region of the field of view and *S* increases further (see Fig.3g-i). Interestingly, transition to the nematic phase during *Drosophila* CE occurs in two steps. The time dependence of *S* for the cells in the dashed rectangle region in Fig. 3b shows a rapid growth from a value of around zero to 0.2 at early times (Fig.3j). Such a process is not found in zebrafish and *Xenopus* experiments, which may be due to a lack of data at short times for these two organisms (see the higher values of S observed in Fig.1j, 2j at t=0). Subsequently, rapid growth starts until the nematic phase is formed, which lasts for about 15 minutes (Fig. 3j). The time dependence of *S* at this late stage is like that in zebrafish and *Xenopus* (see the best fit function, $S(t) = \Sigma - \Gamma\, t^{-\beta}$, shown in Fig.1j, 2j and 3j). The parameter Σ represents the intercept of the curve at t=0, Γ defines the magnitude of S over time, and $\beta$ characterizes the rate of change of S. The nematic order develops before significant changes in cell shape occur (see the inset in Fig.3j), as observed in zebrafish and *Xenopus*. It is striking that all three organisms show similar growth of nematic order despite the different directions of cell orientation, which has no effect as the overall direction is irrelevant in the nematic order.

The decay of $C_S(r)$ (Eq. (2)) in *Xenopus* and *Drosophila* at early times (t<40 or 15 min, see Fig.2j and 3j) is faster compared to zebrafish, which is due to the onset of nematic order with lower values of *S* (< 0.5) in the former two organisms at these times. Strikingly, despite the differences noted above, $C_S(r)$ exhibits long-range spatial correlation, decaying as a power law ($C_S(r) \sim r^{-\alpha}$) at all times for all three organisms (see Fig.1k, 2k, and 3k).

To further explore the spatial and temporal evolution of the nematic phase in *Drosophila*, we calculated the heatmap for *S* at different times (Fig.3l). An ordered nematic phase emerges from the central anterior region, which grows and spreads to the posterior side of the field of view, again like the nucleation and growth process. We conclude that the appearance and growth of a nematic liquid crystal phase, consisting of a few tens to hundreds of cells on a length scale of hundreds of microns, is common to all three organisms.

Finally, to compare the three species, we plotted *S(t)*, and $C_S(r)$, by rescaling of t and distance r for each organism to make them dimensionless, which allows us to compare the spatial/temporal dependent behavior of *S* on equal footing (see Fig.3m-n). This is not only of interest in physics but also hints at a potentially universal behavior in biology as it connects CE across species. Remarkably, regardless of the values of Σ, Γ, and $\beta$ for different species, almost all the data collapsed onto a single master curve (see the solid line in both the figures). It is striking that, despite substantial biological differences in CE between flies and vertebrates[3,34], the analyses of the data show that a similar underlying physical mechanism is operative in all three organisms during the development of the nematic ordered phase in CE. In particular, the algebraic growth of *S(t)*, nucleation and growth mechanism, and long-range spatial correlation in the notochord region are common features in the CE dynamics of the three evolutionarily distant species. We next created a minimal model to explain the universal patterns in the spatiotemporal dynamics leading to the nematic ordered phase in zebrafish, *Xenopus*, and *Drosophila*.

**Theoretical model**

We created a minimal model (see Materials and Methods) to explain the following common phase transition behavior in all three species: (i) Nematic order in the tissue forms by the creation of a domain that grows with time by a process that is reminiscent of nucleation and growth. (ii) *S(t)* increases rapidly at early times and is followed by a slower increase until steady state is reached. The time dependent increase of *S(t)* is well fit by a power law, *S(t)* = $\Sigma$-$\Gamma t^{-\beta}$. (iii) In the notochord region there is a long-range spatial correlation, $C_S(r) \sim r^{-\alpha}$ with $\alpha$ being less than unity.

We first devised a model with only near neighbor interactions in which the orientation of a cell is influenced by only four nearest neighbors (typically involved in T1 transitions) in a two-dimensional lattice. Such a short-range interaction (with the amplitude described by $\mathcal{A}$ in Eqn. (4)) could arise from cell-cell adhesive interactions, for example. Not unexpectedly, we found that while the near neighbor correlation does induce local order, it does not propagate beyond a short distance as time progresses (Fig.4a-c). As a result, the nematic order parameter of the whole tissue is on average approximately zero (see Fig.4d). Thus, the near neighbor interaction model cannot explain the first finding listed above.

Next, we generated a model in which each cell is aligned in the same direction to simulate the effect of global cell alignment (with an amplitude of $\mathcal{B}$ shown in Eqn. (5)), which could be induced by planar cell polarity proteins, for example[35]. With this modification, we find that all the cells in the tissue eventually align globally in the same direction, forming a near-perfect nematic phase (Fig.4e-g). However, the formation and expansion of a small domain at early times that grows with time, which is found in all three species, cannot be explained. This is because each cell changes its orientation independently due to the absence of local interactions among cells in the model (see also Fig.4e-f). In addition, the temporal evolution of *S(t)* grows linearly with time, which contradicts second finding listed above (compare Fig.1j and Fig.4h).

We then considered a third variant, which includes both local and global cell alignment (see Materials and Methods). This model recapitulated the growth behavior of nematic phases found in our experiments (see Fig.4i-k). Moreover, the simulations also predict the presence of defects (disclination lines) as the tissue evolves, which is found in the experiments (see the blue lines in the nematic ordered region in Fig.1-3(g)). The temporal evolution of *S(t)* (see Fig.4l) also shows a rapid growth behavior, and the time-dependent growth is non-linear, *S(t)* = $\Sigma$-$\Gamma t^{-\beta}$, ($\beta$ is larger than zero), both of which are consistent with experiments. Finally, the calculated spatial correlation function $C_S(r)$ shows a similar power-law relation as found in experiments (the inset in Fig.4l and Supp. Fig.S11). Therefore, a two-parameter minimal model with both local (arising from short range cell-cell interactions) and global (arising from long-range patterning) alignment captures the three major findings that are found during CE in all three organisms.

**Loss of Planar cell polarity (PCP) inhibits the long-range order of the nematic phase in *Xenopus.***

The simplicity of the theoretical model allows us to predict the consequences of disrupting the emergence of orientational order during CE. We first considered PCP, as this protein network is required for CE in vertebrates, confers long-range directional information across cellular sheets, and disruptions of PCP cause orientational defects across numerous tissue types and species[32]. Here we knocked down the core PCP protein Prickle-2 (PK2) which is required for *Xenopus* CE (Fig. 5a-b, Video VI)[36,37].

Our theoretical prediction in Fig. 4a-d shows that we should find only the local nematic order in the PCP-mutant tissue, but the global order would be abolished after the long-range interaction is diminished by the knockdown. Satisfyingly, this is exactly what is observed in the knockdown experiments (Fig. 5c-h). Neither SI nor S increase with time as found in the wild-type tissue but undergo small fluctuations as a function of time (Fig. 5i-j). Notably, the value of S is always small (Fig. 5j), lending credence to the notion that the PCP knockdown disrupts long-range interactions. We conclude that the theory correctly predicts inhibition the nematic phase, explaining the results of the wild-type and the knockdowns, thus providing further support for tenets of the theory.

**Loss of adhesion inhibits the short-range order of the nematic phase in *Xenopus.***

Next, we investigated the effect of disrupting short-range interactions by reducing cell-cell adhesion in *Xenopus*. To this end we knocked down Cdh3 (C-cadherin, P-cadherin in mammals), the classical cadherin expressed in the frog embryo, which is required for CE[29,38]. Upon knockdown of Cdh3 the cells are rounded (with smaller *SI* values, see Supp. Fig.12) compared to the wild type cells (see Fig.6a-b). There is no significant change in the cell shape over the course of the experiments (see Fig.6c and Video VII). In addition, the nematic order parameter *S* for cells in the field of view does not show significant changes over time, except for fluctuations around zero (Fig.6d). Thus, loss of Cdh3 inhibits the formation of the nematic phase.

The values of *S* along different directions at different times also show only minor fluctuations around zero (Fig.6e-g, h-j). These data not only show that Cdh3 is required for cell anisotropy but also that short-range interactions are an essential parameter for nematic phase formation in vivo.

To further support this conjecture, we performed mosaic Cdh3 knockdown experiments (Fig.7a-b), where cadherin is knocked down in only a fraction of cells. Remarkably, the mutant and wild-type tissues clearly segregate without interfering with each other (Fig.7c-d). This finding cannot be explained if cadherin also plays an essential role in long-range interactions. The order is completely lost in the knockdown cells (left panel in Figs.7d, f), while it remains in the wild-type cells (right panel in Figs.7d, f), even though the two types of cells are in close contact with each other. Little change of S is found along the anteroposterior axis (Fig.7e). In addition, both SI and S show little changes in the knockdown cells (Fig.7g-i), while they increase with time in the wild-type cells (Fig.7h-j) as observed in Fig.2c, j. Therefore, we believe that cadherin mainly plays a role in driving short-range interactions during the formation of the nematic phase in CE.

**The spadetail mutant (spt/Tbx16) results in a collapse of the long-range order of the nematic phase in zebrafish.**

We further tested the validity of the nematic order framework by disrupting another known molecular regulator of CE with an unclear relationship to short- or long-range interactions. *Spadetail* (*spt*) encodes the transcription factor Tbx16, which is expressed in the cells (paraxial mesoderm) next to the notochord (axial mesoderm) but not in the notochord itself, and it controls boundary formation between the notochord and the paraxial mesoderm[39-41]. While the molecular mechanisms of PCP and Cdh3 are well established the activity of Spt is less clear. Thus, this experiment seeks to determine if the nematic order framework can provide insight into how Spt controls convergent extension.

When comparing representative snapshots of the spt-mutant and wild-type zebrafish tissues (see Fig.8a-b), the mutant tissue lacks notochord boundaries as expected[41]. In the region where we expected to see the anisotropic notochord cells, we observed a stripe of highly constricted cells (see the dashed rectangle region in Fig. 8a, and Video VIII). In addition, the time dependence of the cell shape, *SI(t)*, and order parameter, *S(t)*, of cells in the mutant tissue change non-monotonically with time (see Fig.8c-d). Initially, *SI* increases linearly with time and there is a rapid growth of *S* indicating the formation of a nematic phase. Unlike the control, however, after a short time, both *SI* and *S* decrease rapidly. Although many cells initially align in the mediolateral direction, especially early in the time course (Fig.8e), there are many disordered regions scattered throughout the tissue. Thus, *Spt* tissue fails to display the near-perfect nematic order found in wild-type tissue (Fig.1g). Interestingly, it is similar to our simulations that lack global cell alignment (see Fig.4a-d).

At later times, the ordered region in the mutant also decreases (Fig.8h), leading to the decay of S(t) (Fig.8d). In addition, in the wild type, the order parameter *S* has a peak in the presumptive notochord region (Fig.1f). Such a peak is absent in the spt mutant, either along the mediolateral or anteroposterior directions at early or late times (Fig.8f-g, i-j). Although the spt-mutant tissue shows some degree of nematic order in certain regions, it is substantially weakened compared to the notochord region in wild-type. These findings show that disruption of global patterning, but not local cell interactions, alters the formation and maintenance of the nematic order phase. Further, these data indicate a time dependence for the spt mutant as nematic order is not abolished but instead collapses during notochord formation. We suspect that the collapse relates to the timing of boundary formation and can conclude that spt results in a delayed collapse of nematic order after the onset of convergent extension.

**Evaluating the mutants with the theoretical model.**

Finally, we further assessed the validity of our model, by asking if changing parameters can evoke outcomes that reflecting those seen for Cdh3 and *spt* experiments. First, we modelled the results of the Cdh3 knockdown in *Xenopus* CE. We expect that cell-cell adhesion induced by Cdh3 would influence the local cell alignment. By changing the parameter for the local cell interaction, parameter $\mathcal{A}$ in Eq.(7) to a much smaller value (2500 times smaller) than that of the wild-type (see Supp. Table I), with a mild reduction of the global parameter $\mathcal{B}$ in the same equation (one third of the value of wild-type), our model gives results similar to those observed in experiments (see Fig.9a-d). Therefore, our model suggests that the disruption of the nematic phase caused by the knockdown of Cdh3 in *Xenopus* results primarily due to the loss of short-range interactions.

We next modeled the effect of loss of spt by changing in the global parameter, expecting that this will allow the model to recapitulate the pattern of nematic order observed in the mutants. We therefore changed the sign of the global parameter $\mathcal{B}$ in our model (Model (iii) in Materials and Methods) after t = 1,000 time steps, see Fig.9e-h), which has the effect of disrupting the cell alignment along the horizontal axis due to the potential change in the global alignment the force field, fluctuations of cell orientation, or other effects after the spt loss. We observed an increase in the ordered phases for a period after switching this parameter. As the tissue evolves, it starts to become more disordered and the order parameter $S$ for the whole tissue also decreases (see Fig.9g-h), which is consistent with experimental findings in Fig.8d. Thus, by changing the value of the global order parameter, we can rationalize collapse of the nematic phase caused by the mutation of the *spt* gene.

Taken together, the minimal two parameter theoretical model with both local and global cell alignment, explains *all* the measured salient features of the nematic phase formation in gastrulation of zebrafish, *Xenopus* and *Drosophila*. The prediction for the effect of the global cell alignment in our theoretical model is validated by the PCP-mutant experiments. Moreover, other mutant experiments involving perturbation of local and global cell alignment can also be explained by the theoretical model using suitable values for the two parameters. We should note that despite the success of the model, we have not established a connection between the two parameters and the molecular processes that drive CE. Nevertheless, the combined experimental and theoretical study suggests that future experiments that are designed to systematically interfere with the CE process in all three or more organisms can be used to further support the conclusions of this study.

**Effect of noise.**

Despite the success of the two-parameter model, fluctuations in the nematic phase (see Figs. 1l, 2l and 3l), where an ordered region could become disordered at later times are not considered. To capture the effect of fluctuations, we added an extra noise term C $\zeta$(t) to the model (iii), where the noise strength C is a constant and $\zeta$(t) is white noise. One example for the cell orientation at different times using the new model is shown in Fig. 10(a) and the time dependent changes in S for all cells is shown in Fig. 10(c).

In the presence of noise, *S(t)* continues to exhibit a similar power-law behavior. However, the value of *S(t)* fluctuates depends on the trajectory (see the curves in gray), indicating that the ordered region occasionally becomes disordered (see the inset in Fig. 10(c)). An example of such a fluctuation of ordered region is shown in Fig. 10(a) (see the region enclosed by the black box) and Fig. 10(b). We surmise that the presence of noise does not influence the major results of this work. Therefore, a simple two-parameter model suffices to capture the nematic phase transition during CE process.

# Discussion

We found that a nematic order phase arises during gastrulation in zebrafish, *Xenopus*, and *Drosophila*. All three organisms show that the nematic phase forms during CE by a nucleation and growth mechanism. The preferred orientational ordering of cells is accompanied by a slow power-law decrease associated with the spatial correlation in the nematic order parameter, suggestive of long-range order. These findings can be nearly quantitatively explained by a simple theoretical model and demonstrate that the underlying tissue-level mechanism is conserved.

We find it surprising that the emergence of the transition to the nematic phase occurs through a common nucleation and propagation mechanism in each animal. However, there were interesting differences. For example, the orientation and time dependence of cellular anisotropy was strikingly different in *Drosophila* compared to the vertebrates. The properties of the nematic phase of the tissues allow them to flow like a liquid while maintaining orientational order[42], which could be essential for their biological functions or may make tissues more robust to environmental perturbations. It is tempting to suggest the phase transition to the nematic state may be universal feature of animal embryogenesis, which implies that the cellular process of convergent extension may be conserved from flatworms to mammals[2]. To establish that the predicted phase transition is universal requires additional studies involving other animals.

Interestingly, we also found defects in the nematic phase of tissues. Such defects have a role in cellular processes such as cell death and extrusion and cytoskeletal organization[43]. It will be interesting to investigate the relationship between these defects and mechanical properties of "active liquid crystals"[44-47]. Thus, additional studies are needed to search for liquid crystalline phases in biology at different scales and explore their possible biological significance[48-52].

In summary, despite vast differences, the three organisms studied attain the same physical state, with the properties of a nematic liquid crystal, during CE. We believe it is unexpected, especially given that three examples considered here are evolutionarily and morphologically distinct. Nevertheless, they reach the same physical state during CE by common mechanism. Surprisingly, the results of all the experiments are quantitatively explained by a single theoretical model with just two parameters. Our work also adds to the growing idea of applying liquid crystal physics to several biological problems involving cell collectives[43]. The application to tissue scale cell movements that drive convergent extension has not been previously explored. We believe that such a phase transition from a disordered to ordered state must have an important functional role during embryogenesis.

# Methods:

All animal research performed for this study was approved by IACUC.

**Image analysis:** A single z-plane through live zebrafish, *Xenopus*, or *Drosophila* embryos expressing membrane EGFP (or mCherry), was selected for image analysis[19]. We first used "Cellpose", a deep learning-based segmentation method[53], to detect cell boundaries from snapshots of movies during zebrafish, *Xenopus* and *Drosophila* gastrulation (see Fig. 1(a)-(b), Fig. 2(a)-(b) and Fig. 3(a)-(b) for example). We then changed the image segmentation masks into gray scale images, which we used to track cells through ImageJ TrackMate plugin[54]. Finally, the position, area, perimeter, and the major and minor axes of cells are exported from TrackMate plugin for further analysis.

**Statistics and Reproducibility**

The wild-type experiments for zebrafish and *Drosophila* were repeated three times, and the experiments for *Drosophila* were repeated twice. Each of the *Xenopus* perturbations (PCP-protein knockdown, C-cadherin knockdown, C-cadherin mosaic mutant) has been repeated more than three times. We repeated the experiment twice for the spt mutant in zebrafish. The results are similar to the example shown in the figures.

**Two-fold orientational order parameter $\psi_2$:** The n-fold orientational order parameter $\psi_n$ is defined by[55],

$$\psi_n(i) = \frac{1}{\sum_j l_{ij}^2} \sum_{j \in N(i)} l_{ij}^2 e^{in\theta_{ij}} \quad , (3)$$

where the sum is taken over all the nearest neighbors of the i$^{th}$ cell, $l_{ij}$ is the length of the edge shared between the Voronoi cells i and j, and $\theta_{ij}$ is the angle between the vector pointing from cell i to j, and mediolateral axis (see the inset in Supp. Fig. 5(a)). Here, we consider n=2, the 2-fold orientational order parameter $\psi_2$, shown in Supp. Fig. 5(e)-(f).

**Models for the nematic phase during CE:** To rationalize the experimental findings, we introduce a simple two-dimensional lattice XY-type model. Each cell is located on a lattice site (i, j) (see Fig. 4(a)). The cells are allowed to change their orientation at each time step. Three variants of the model are considered, depending on the rules that govern how cells change their orientation.

**Model (i)**: The cell orientation cell, $\theta_i$ (the angle between the long axis of a cell and the horizontal axis), is influenced only by the orientation of four nearest (up, down, left, and right) neighbors. The temporal evolution of $\theta_i$ is described by,

$$\frac{d\theta_i}{dt} = -\mathcal{A}\Sigma_j \sin(\theta_i - \theta_j), \quad (4)$$

where the summation is taken over the four nearest neighbors of cell $i$, and $\mathcal{A}$ is a constant. That quartet of cells exchange neighbors via T1 transitions during CE is well known[3,6] (see Supp. Fig.8). More recently[56], vertex models have been used to describe this process theoretically, with focus on *Drosophila*. The goal of Model (i) is to assess the extent of order that arises due to short range cell-cell interactions involving a quartet of cell, without considering T1 processes explicitly.

**Model (ii):** Cell orientation, $\theta_i$, is only influenced by a global force field, leading to,

$$\frac{d\theta_i}{dt} = -\mathcal{B}\theta_i, \quad (5)$$

where $\mathcal{B}$ is a constant. In this variant all the angles $\theta_i$ evolve independently. In other words, there is complete absence of cooperativity, and the dynamics is controlled by the magnitude of $\mathcal{B}$.

**Model (iii):** In the third variant, the dynamic of $\theta_i$, is influenced by both the local cell ($\mathcal{A}$) alignment and the global ($\mathcal{B}$) fields, which leads to,

$$\frac{d\theta_i}{dt} = -\mathcal{A}\Sigma_j \sin(\theta_i - \theta_j) - \mathcal{B}\theta_i. \quad (6)$$

The Hamiltonian, $\mathcal{H}$, of the tissue is described by,

$$\mathcal{H} = \Sigma_i(-\mathcal{A}\Sigma_j \cos(\theta_i - \theta_j) + \frac{1}{2}\mathcal{B}\theta_i^2), \quad (7)$$

The models (i) and (ii) are special cases of Eq. (7). The first term in the above equation is taken from the two-dimensional X-Y model[57], and the second term is similar as the energy term in the mean field Maier–Saupe theory[58]. To simulate the temporal evolution of each cell orientation using Eqs. (4)-(6), a uniform distribution for $\theta_i$ $(-\pi/2, \pi/2)$ is applied initially. We used periodic boundary conditions in the simulations. A lattice size of 20×20 is used in the main text. Different lattice sizes (30×30, 50×50) are considered to check the finite size effect (see Supp. Fig.13-14). Almost the same results are found for the three models under different lattice sizes, indicating the absence of the finite size effect. To consider the effect of fluctuations, we added an extra noise term, C $\zeta$(t), to the model in Eq. (6), where C is a constant and $\zeta$(t) is white noise with $\langle\zeta(t)\zeta(t')\rangle = \delta(t - t')$.

**Zebrafish embryo manipulations, injections, and imaging:**

*Zebrafish strains and embryo staging*: Adult zebrafish were raised and maintained according to established methods[59] in compliance with standards established by the Washington University Animal Care and Use Committee. Embryos were obtained from natural matings and staged according to morphology as described[60]. All WT studies were carried out in animals of the AB background. Additional lines used include *spt*$^{m423}$ [61]. Embryos of these strains generated from heterozygous intercrosses were genotyped by PCR after completion of each experiment.

*Microinjection of zebrafish embryos:* One-celled embryos were aligned within agarose troughs generated using custom-made plastic molds and injected with 1-3 pL volumes using pulled glass needles. Synthetic mRNAs for injection were made by *in vitro* transcription from linearized plasmid DNA templates using Invitrogen mMessage mMachine kits. 100 pg *membrane Cherry* (a kind gift from Dr. Fang Lin), or 50 pg membrane *eGFP* (*7*) mRNA was injected per embryo.

*Microscopy:* Live embryos expressing fluorescent proteins were mounted in 0.75% low-melt agarose in glass bottomed 35-mm petri dishes for imaging using a modified Olympus IX81 inverted spinning disc confocal microscope equipped with Voltran and Cobolt steady-state lasers and a Hamamatsu ImagEM EM CCD digital camera. For time-lapse series, 60 μm z-stacks with a 2 μm step, were collected every three or five minutes for three or four hours using a 40x dry objective lens. Embryo temperature was maintained at 28.5°C during imaging using a Live Cell Instrument stage heater. When necessary, embryos were extracted from agarose after imaging for genotyping.

***Xenopus* embryo manipulations, injections, and imaging:** Embryos were acquired through in vitro fertilization. Female *Xenopus* were injected with 600 units of human chorionic gonadotropin and incubated overnight at 16$^{o}$C. Eggs were then squeezed from the female *Xenopus* and fertilized. Eggs were dejellied two hours after fertilization using 3% cysteine (pH 8) and washed and reared in 1/3X Marc's Modified Ringer's (MMR) solution. Embryos were placed in 2% ficoll in 1/3X MMR for microinjections and then returned to 1/3X MMR 30 minutes after injections. A Parker's Picospritizer III and an MK1 manipulator were used for microinjections. Four-cell embryos were injected in the dorsal blastomeres to target the presumptive dorsal marginal zone. Membrane-RFP mRNA was injected at a concentration of 100pg per blastomere. Cdh3 morpholino was injected at a concentration of 10ng per blastomere. For mosaic Cdh3 knockdown a single dorsal blastomere was injected with 10ng of Cdh3 morpholino and 100pg of membrane GFP mRNA and both blastomeres were injected with membrane RFP mRNA. Prickle2 morpholino was injected at 25ng per blastomere. For dissections, stage 10.25 embryos were moved to Danilchik's for Amy (DFA) medium and Keller explants were excised using eyelash hair tools. Explants were maintained in DFA following dissection and time-lapse movies were collected ~5 hours after dissection. The explants take several hours to begin CE after dissection from the embryo. Once the explants begin CE, it closely follows what we observe in the embryo. The time-lapse images in this work show the earliest steps of CE. Images were acquired using a Nikon A1R microscope with a two-minute time interval and at a z-depth of ~5μm into the explant (above the superficial surface/coverslip). We exclusively analyzed the axial mesoderm because the notochord region is very broad along the mediolateral axis at this stage and filled the entire field of view at the magnification required for our cell scale analysis.

***Drosophila* embryo movies:** We chose the Drosophila germ band (lateral epidermal progenitors) because this is the cell population that converges and extends during Drosophila gastrulation[31]. The movies analyzed were collected and published in Sawyer et al. 2011[62]. Wildtype Drosophila embryos expressing DEcadherin-GFP under control of the ubiquitin promotor and myosin light chain–mCherry (= Spaghetti Squash [sqh]) were filmed during stage 7 of embryonic development. Live imaging was performed with a PerkinElmer (Waltham, MA) UltraView spinning disk confocal ORCA-ER camera, Nikon (Melville, NY) 60× Plan Apo NA 1.4 or 100× Plan ApoVC NA 1.4 objectives, and MetaMorph software (Molecular Devices, Sunnyvale, CA).

# Data availability:

Videos and source data are provided with this paper.

# Code availability:

Code for the model is available from a GitHub repository:
[https://github.com/xinlee0/nematic_-embryos].

# Acknowledgements:

This work is supported by the National Science Foundation (grant no. PHY 2310639) (D.T.), the Collie-Welch Chair through the Welch Foundation (F-0019) (D.T.), NICHD (grant no. R01HD099191) (J.B.W.), and NIH R35 (grant no. GM118096) (M.P.).

# Author Contributions:

X.L., R.J.H., J.B.W, and D.T. conceived the study. X.L. and R.J.H. performed experiments and analysis. M.W. performed the zebrafish experiments. J.S. and M.P. performed the *Drosophila* experiments. X.L., R.J.H., J.B.W, and D.T. wrote the initial manuscript. X.L., R.J.H., J.B.W, D.T.,

M.P. and M.W. edited the manuscript. All the authors have seen and approved the final version of the manuscript.

# Competing Interests:

No competing interests declared.

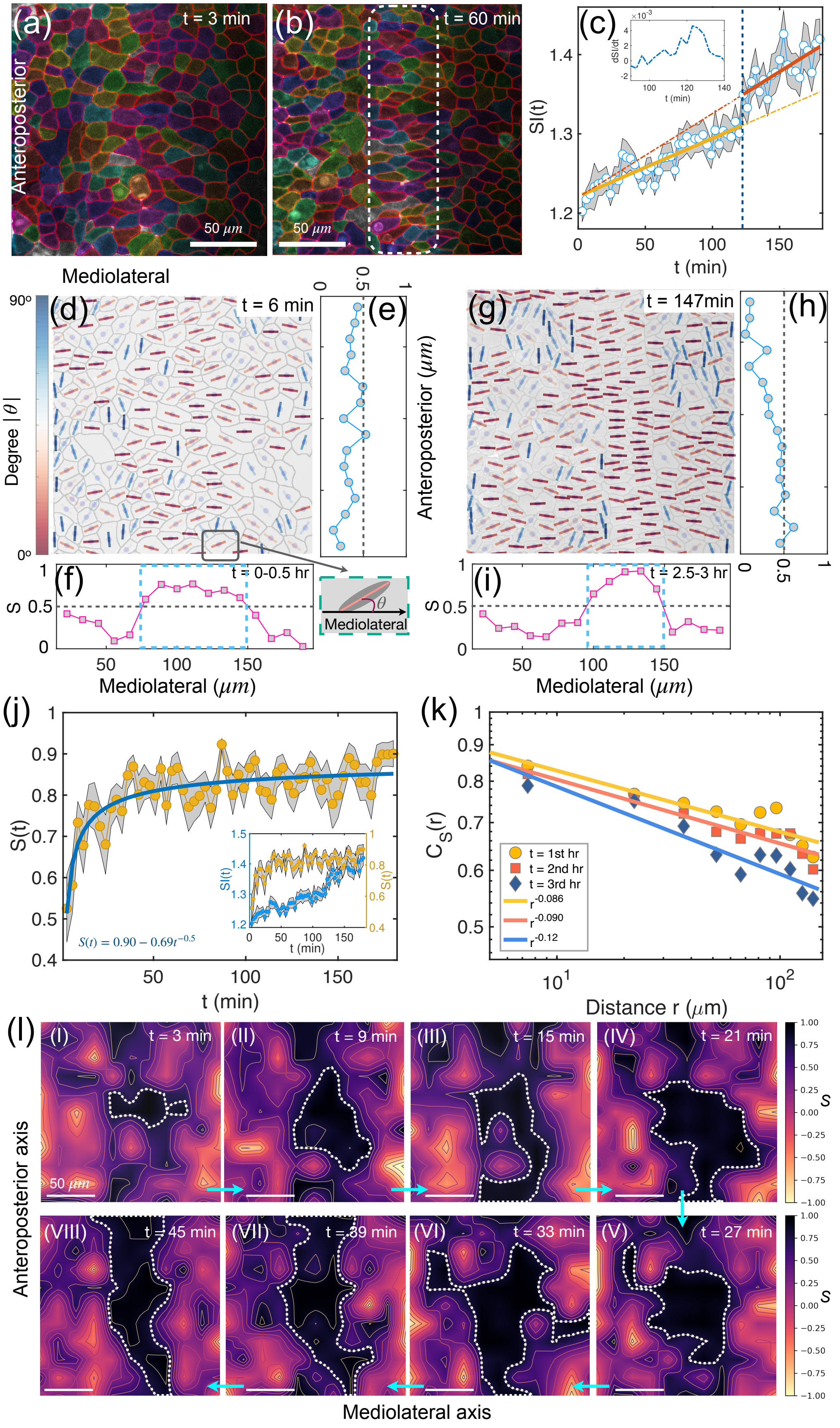
(a)
t = 3 min
Anteroposterior
50 μm
(b)
t = 60 min
50 μm
Mediolateral
(c)
SI(t)
t (min)
dSI/dt
(d)
t = 6 min
90°
0°
Degree |θ|
(e)
(f)
t = 0-0.5 hr
S
Mediolateral (μm)
Mediolateral
θ
(g)
t = 147min
(h)
Anteroposterior (μm)
(i)
t = 2.5-3 hr
Mediolateral (μm)
(j)
S(t)
t (min)
S(t) = 0.90 − 0.69t^−0.5
SI(t)
(k)
C_S(r)
t = 1st hr
t = 2nd hr
t = 3rd hr
r^-0.086
r^-0.090
r^-0.12
Distance r (μm)
(l)
(I) t = 3 min
(II) t = 9 min
(III) t = 15 min
(IV) t = 21 min
(VIII) t = 45 min
(VII) t = 39 min
(VI) t = 33 min
(V) t = 27 min
50 μm
Anteroposterior axis
Mediolateral axis
S

Figure 1. **Nematic order during zebrafish convergent extension (CE).** (a)-(b) Snapshots of zebrafish tissue at different times. The dashed rectangle in (b) shows the notochord location. The color associated with each cell is for illustration. (c) Temporal evolution of the shape index *SI(t)* of cells in the notochord region shown in (b). The orange and red lines are linear fits ($SI(t)=7\times10^{-4}t+1.22$, $10^{-3}t+1.22$) with open circles (mean values), separated by the blue dashed line, where a jump in *SI(t)* occurs (see the inset for the derivative of *SI(t)*). The shaded area indicates the standard error of mean (SEM). n≈60 cells from Video I. (d) Cell orientation, defined by the angle, $\theta$, between the long axis of cells (see the short lines) and the horizontal (mediolateral) axis of the embryo, (the inset at the bottom right of (d)) at t=6 min. The short lines are color coded by the $|\theta|$ value (a key is shown to the left of panel d). (e)-(f) The nematic order parameter, *S*, as a function of the cell position along the anteroposterior or mediolateral axis. Each curve is averaged over 10-time frames spanning 30 minutes. (g)-(i) Same as (d-f) except at later timepoints of CE. (j) Time dependent changes in *S(t)* of cells in the notochord region identified in (b). The solid line is a power-law fit ($S(t)=0.9-0.69t^{-0.53}$) with circles being the mean values. The shaded area shows the SEM. The inset shows the *SI(t)* and *S(t)* in the same plot. (k) The spatial correlation, $C_S(r)$ (Eq. (2)), of cells in the notochord region at different times. The solid lines show a power-law decay. The functional forms of the decay are displayed in the figure. (l) (I)-(VIII) The spatial-temporal evolution of *S* shows the propagation of the nematic order (see the black region enclosed by the white dotted line where *S*>0.8). Each figure in (I)-(VIII) is averaged over two successive time frames (3-minutes interval). The scale bar in (a), (b) and (l) is 50 μm.

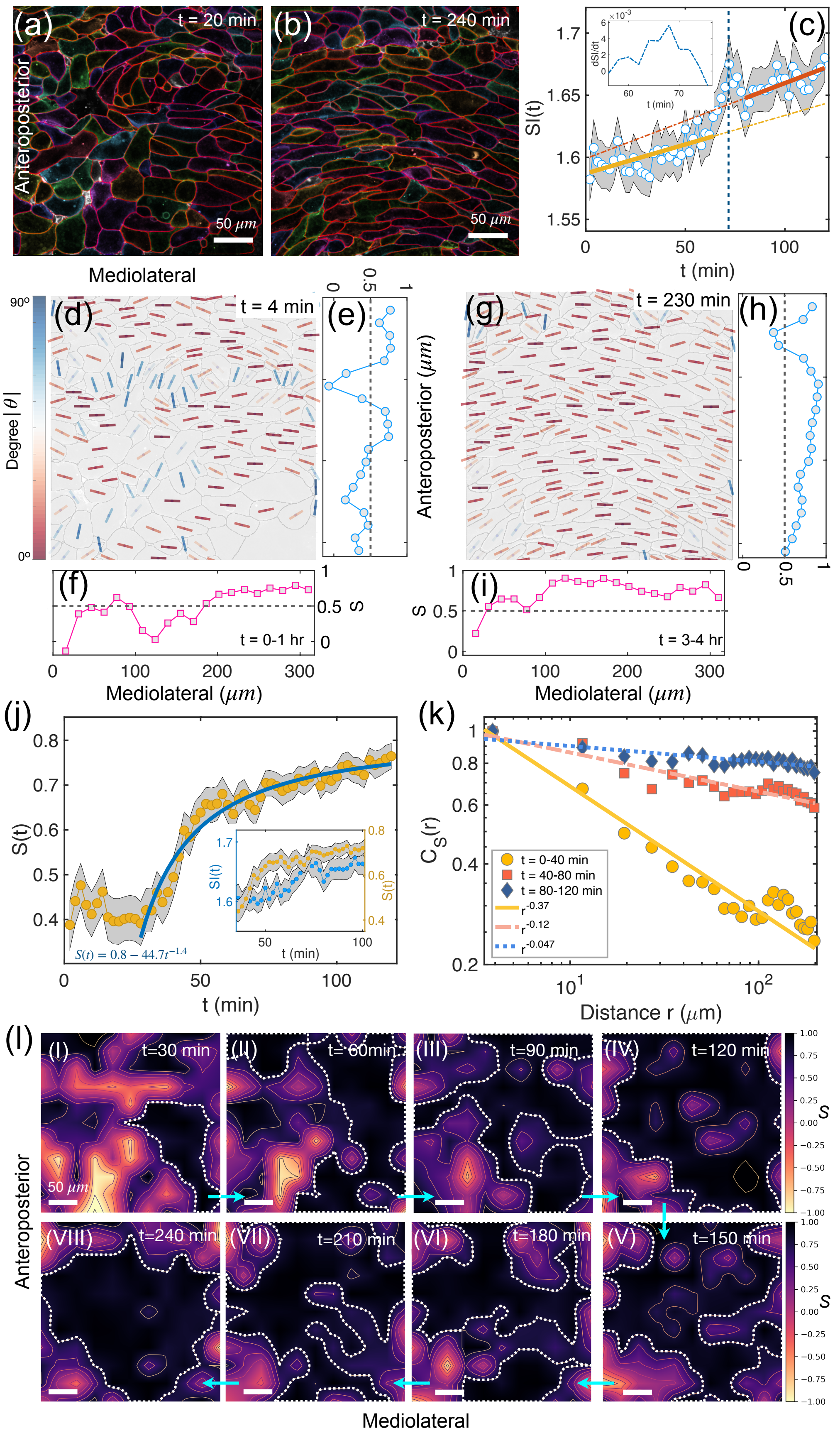

Figure 2. **Emergence of a nematic phase during *Xenopus* CE.** (a)-(b) Snapshots of *Xenopus laevis* tissue at different timepoints. (c) Time dependent changes in the shape index *SI(t)* of the notochord cells in the field of view. The orange and red lines are linear fits (*SI(t)*=$4.7\times10^{-4}$t+1.59, $6.1\times10^{-4}$t+1.6) with open circles (mean values). The data at early and late timepoints are separated by a jump (blue dashed line, see also the inset for the derivative of *SI(t)*) in *SI(t)* at t=70 min. The open circles give the mean *SI* values of all cells in the field of view at each time point. The shaded area shows the SEM. n≈200 cells from Video IV. (d) Cell orientation in Xenopus tissues early during CE. The short lines are color coded by the value of $|\theta|$ (defined in Fig. 1d.) (e)-(f) Nematic order parameter, *S*, as a function of the cell position along anteroposterior or mediolateral axis at an early timepoint. The two curves are obtained by averaging over 30 successive time frames spanning one hour. (g)-(i) Same as (d)-(f), except later. (j) Temporal evolution of the nematic order parameter *S* of cells in the field of view. The solid line is a power-law fit (functions listed in the figure) with circles (mean values). The shaded area shows the SEM. The inset shows the *SI(t)* and *S(t)* in the same plot. (k) Spatial correlation, *S*, of notochord cells at different times. Solid lines are a power-law fits (functions listed in the figure) of the data at different times. (l) (I)-(VIII) Same as Fig.1(l), showing the propagation of the nematic order (see the black region enclosed by the white dotted line). Each figure in (I)-(VIII) is averaged over three successive time frames (2-minutes interval). The scale bar in (a), (b) and (l) is 50 μm.

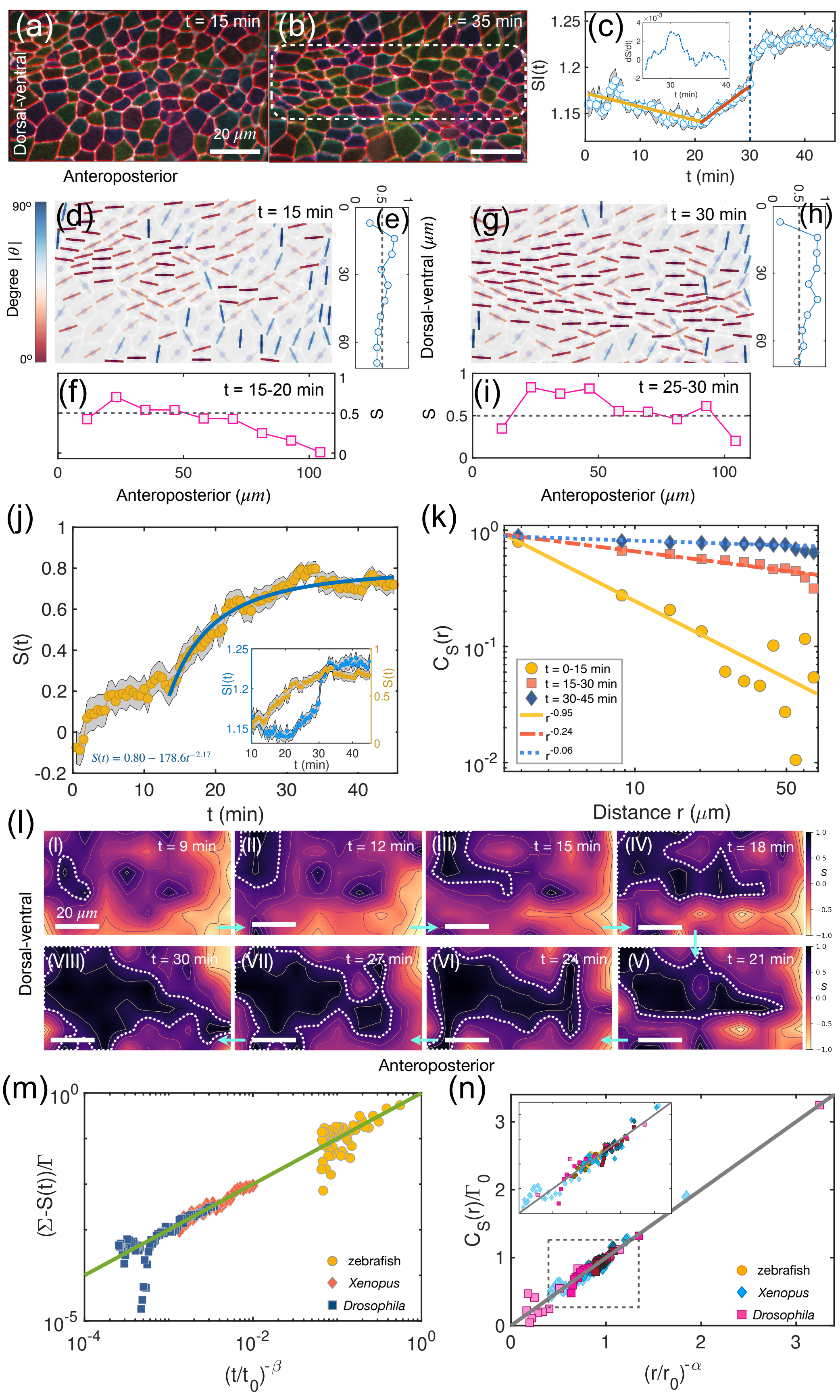
(a)
t = 15 min
Dorsal-ventral
20 μm
Anteroposterior
(b)
t = 35 min
(c)
SI(t)
t (min)
dS/dt
(d)
t = 15 min
Degree |θ|
90°
0°
(e)
(f)
t = 15-20 min
Anteroposterior (μm)
(g)
t = 30 min
(h)
Dorsal-ventral (μm)
(i)
t = 25-30 min
(j)
S(t)
t (min)
S(t) = 0.80 − 178.6t^−2.17
(k)
C_S(r)
Distance r (μm)
t = 0-15 min
t = 15-30 min
t = 30-45 min
r^-0.95
r^-0.24
r^-0.06
(l)
t = 9 min
t = 12 min
t = 15 min
t = 18 min
t = 21 min
t = 24 min
t = 27 min
t = 30 min
Dorsal-ventral
Anteroposterior
(m)
(Σ-S(t))/Γ
(t/t0)^-β
zebrafish
Xenopus
Drosophila
(n)
C_S(r)/Γ0
(r/r0)^-α

Figure 3. **Emergence of nematic order in *Drosophila* during CE.** (a)-(b) Representative snapshots of lateral views of the *Drosophila* germband. (c) Shape index *SI(t)* as a function of time for cells located in the dashed rectangle in (b). Linear fits (*SI(t)*=$-1.5\times10^{-3}$t+1.17, $4.4\times10^{-3}$t+1.05) with open circles (mean values) at different times, shown in orange and red. The shaded area shows the SEM. n≈100 cells from Video V. (d) Cell orientation is defined by the angle, $\theta$, between the long axis of cells (see the short lines) and the anteroposterior axis. The short lines are color coded by the value of $|\theta|$. (e)-(f) *S* as a function of the cell position along the dorsal-ventral, anteroposterior axis. The two curves are obtained by averaging over 10 successive time frames in five minutes. (g)-(i) Same as (d)-(f), except at later times. (j) Temporal evolution of *S* for cells in the dashed rectangle in (b). The solid line is a power-law fit (*S(t)* = Σ-Γ $t^{-\beta}$ , with functions listed in the figure) of the data with circles (mean values). The shaded area shows the SEM. The inset shows the *SI(t)* and *S(t)* in the same plot. (k) Spatial correlation of *S* for cells in the dashed rectangle in (b) at different times. The solid lines are a power-law fit ($C_S(r) \propto r^{-\alpha}$) at different times. (l) (I)-(VIII) Same as Fig.1(l), it shows the propagation of the nematic order during *Drosophila* development (see the black region enclosed by the white dotted line). Each figure in (I)-(VIII) is averaged over three successive time frames. (m) Scaled order parameter *S* as a function of the scaled time $t/t_0$ ($t_0$=1min) for three organisms: zebrafish, *Xenopus*, and *Drosophila*. (n) Same as (m), except for the scaled spatial correlation of *S* over the scaled distance $r/r_0$. $r_0$, mean cell size, is taken by 10μm for zebrafish and *Drosophila*, and 20μm for *Xenopus*. The same symbol but in different colors represents data taken at different times for the same organism. The inset is a zoom-in view of the dashed box. The scale bar in (a), (b) and (l) is 20 μm.

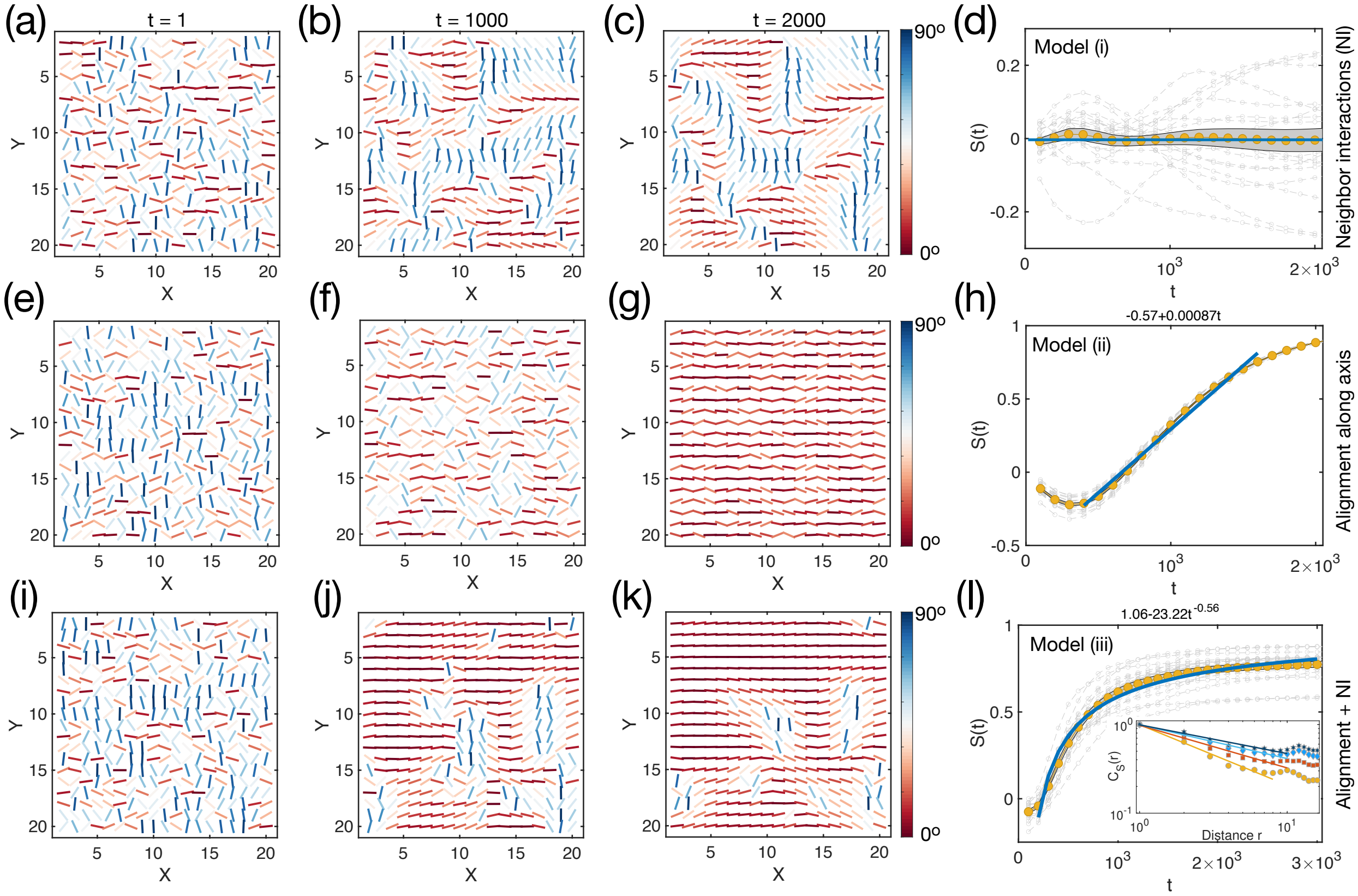

Figure 4. **Models for nematic order formation**. (a)-(d) Results from model (i) (Materials and Methods), where each cell orientation ($\theta$) is influenced only by four neighbors. (a)-(c) Cell orientation at different times t. The parameter $\mathcal{A}$ = 2.5×10^-3^. (d) Time dependent changes in $S$ for all cells in (a). Each curve in gray represents one realization. Data in the brown dot is the mean value averaged over 20 trajectories. The shaded area shows the SEM. The solid line in navy blue corresponds to S = 0. (e)-(h) Results from model (ii), where each cell orientation is influenced only by a global field, leading alignment along the X-axis independently. The navy-blue line in (h) is a linear function, listed at the top of the figure. The parameter $\mathcal{B}$ = 1×10^-3^. (i)-(l) Results from model (iii), Eq. (6). Navy blue line in (l) shows a power-law behavior with the function listed at the top of the figure. The values of $\mathcal{A}$ and $\mathcal{B}$ are 2.5×10^-3^, 1×10^-3^, respectively. The inset in (l) shows the spatial correlation, $C_S(r)$, of cells at different times from simulations. The solid lines show the power-law decay.

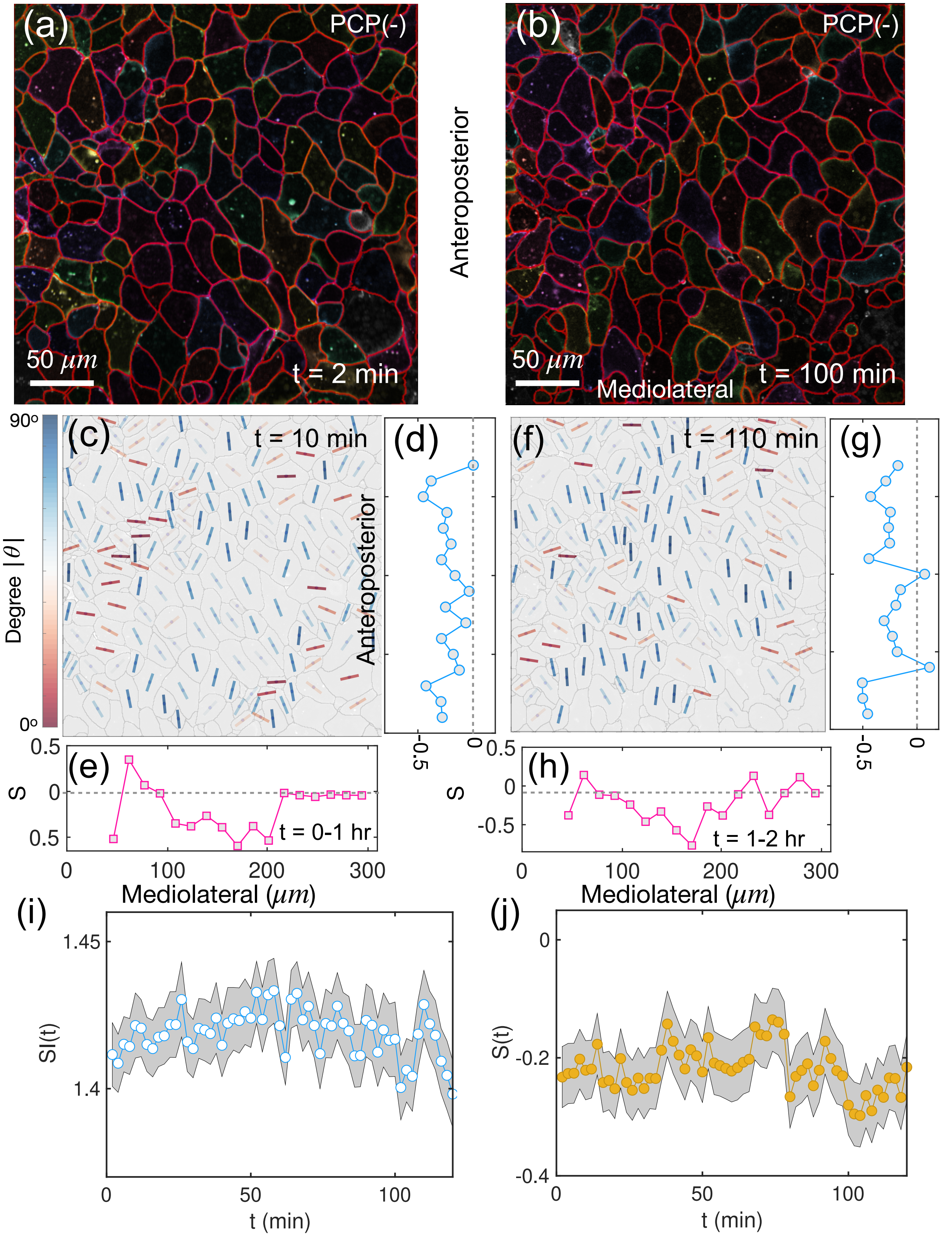

Figure 5. **Reduced nematic order in PCP mutant *Xenopus*.** (a)-(b) Two representative images of PCP-protein knockdown *Xenopus* tissue. (c) Orientation of cells in *Xenopus* tissue (PCP-) at an early timepoint. The short lines are color coded by the value of $|\theta|$. (d)-(e) Nematic order parameter, *S*, as a function of the cell position along the anteroposterior, mediolateral axis. Each curve is averaged over thirty successive time frames spanning 1 hour. (f)-(h) Same as (c)-(e), except at later timepoints. (i), (j) Dependence of *SI(t)* and the nematic order parameter *S(t)* as a function of t in *Xenopus* tissue with PCP protein knockdown. Open circles show the mean values, and the shaded area indicates the SEM. n≈200 cells from Video VI.

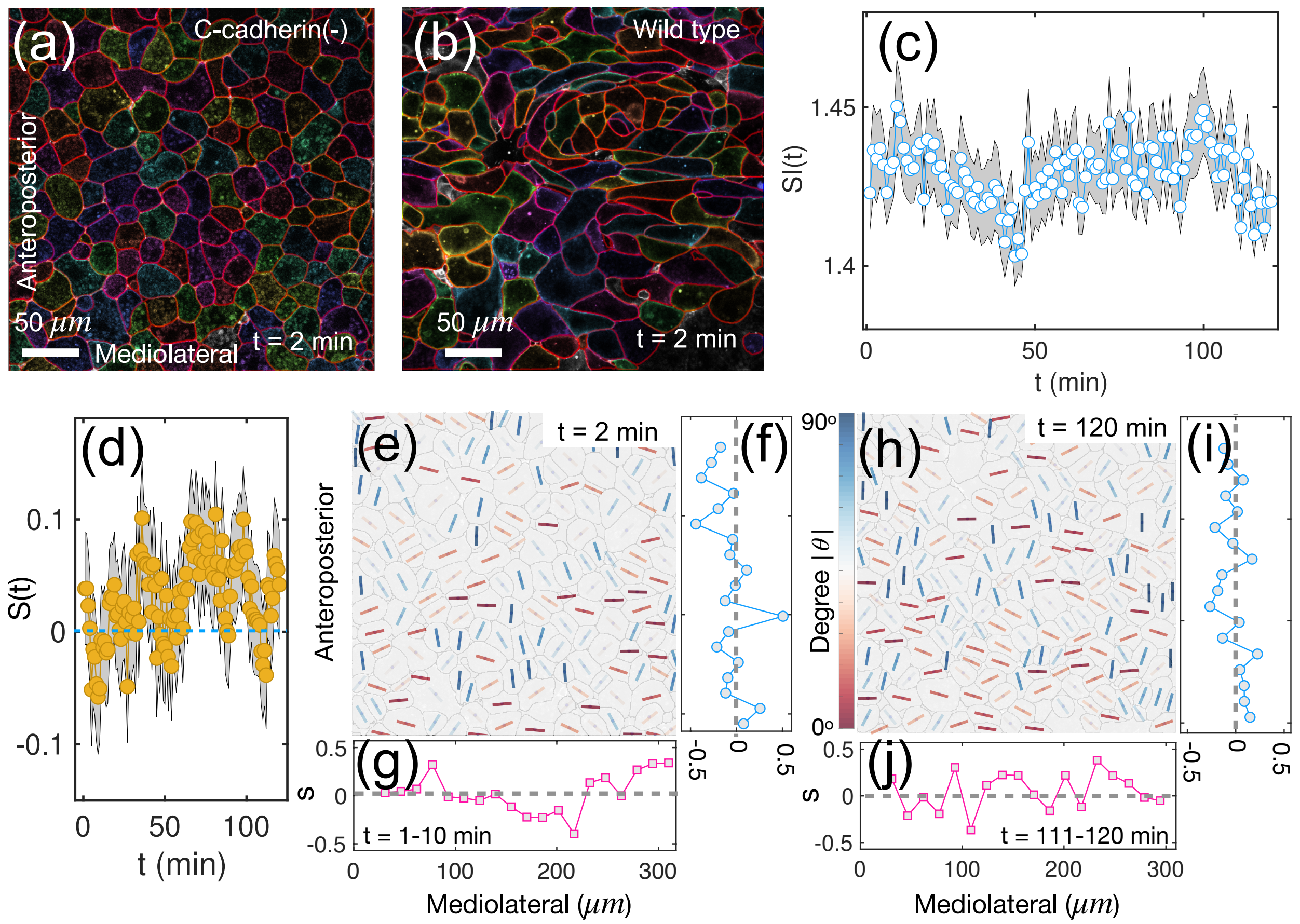

Figure 6. **Reduced nematic order in C-cadherin mutant *Xenopus*.** (a)-(b) Two representative snapshots of C-cadherin knockdown or wild-type *Xenopus* tissue. (c), (d) Dependence of *SI(t)* and the nematic order parameter *S(t)* as a function of t in Xenopus tissue with C-cadherin knockdown. Open circles show the mean values, and the shaded area indicates the SEM. n≈200 cells from Video VII. (e) Orientation of cells in Xenopus tissue (C-cadherin-) at an early timepoint. The short lines are color coded by the value of $|\theta|$. (f)-(g) Nematic order parameter, *S*, as a function of the cell position along the anteroposterior, mediolateral axis. Each curve is averaged over ten successive time frames spanning 10 minutes. (h)-(j) Same as (e)-(g), except at later timepoints.

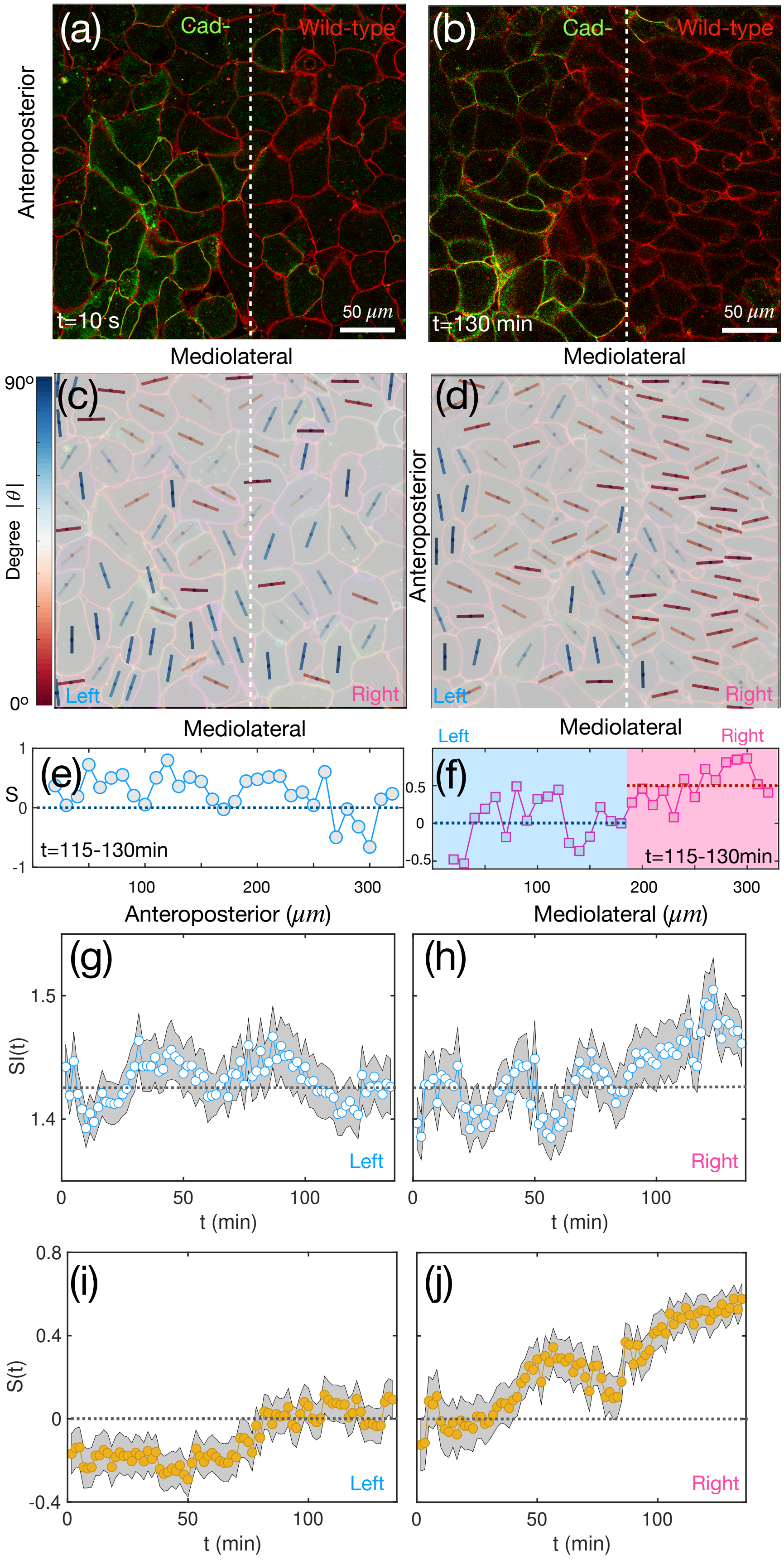

Figure 7. **Effects in C-cadherin mosaic mutants.** (a)-(b) Images of *Xenopus* tissue with C-cadherin mosaic mutant at t=10 s (a) and t = 130 min (b). The cells labeled in green (red) are C-cadherin knock down (wild-type). (c)-(d) Cell orientation corresponds to the two snapshots in (a)-(b). The short lines are color coded by the angle (|θ|) between the long axis of cells and the horizontal axis of the tissue. (e)-(f) The nematic order parameter, *S*, as a function of the cell position along the anteroposterior or mediolateral axis. Each curve is averaged over 10-time frames. A clear jump in the value of *S* is found along the mediolateral axis, where mutant and wild-type cells segregate clearly. (g)-(h) Temporal evolution of the shape index *SI(t)* for cells on the left (mutant-type) or right (wild-type) region of the tissue. (i)-(j) Time dependent changes in *S(t)* of cells on the left or right region as in (g)-(h). The differences between the mutant and the wild type are dramatic. Open circles in (g)-(j) show the mean values, and the shaded area indicates the SEM.

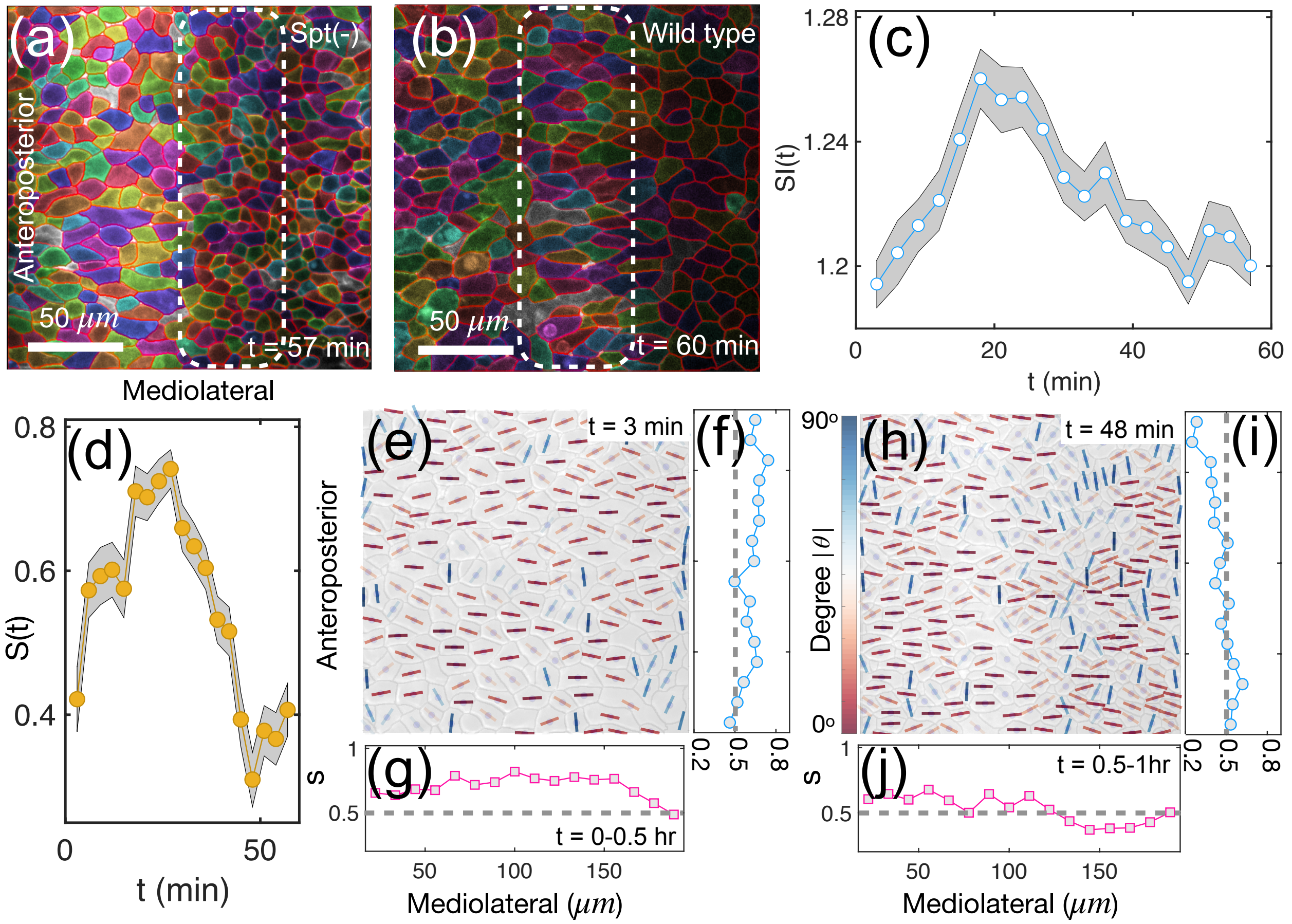

Figure 8. **Collapse of nematic order in spt mutant zebrafish.** (a)-(b) Two representative snapshots of zebrafish tissue in spt-mutant or wild type. The dashed rectangle shows cells around the middle line. (c)-(d) *SI(t)* and *S(t)* as a function of t for the spt-mutant cells in the field of view. The mean (SEM) is shown in open circles (shaded area). n≈300 cells from Video VIII. (e) Cell orientation ($\theta$) in zebrafish spt-mutant early times. (f)-(g) The orientational order parameter, $S$, as a function of the cell position along the anteroposterior, mediolateral axis at early times. Lines are averaged over ten successive time frames spanning 30 minutes. (h)-(j) Same as (e)-(g), except at later times. The scale bar in (b) and (l) is 50 μm.

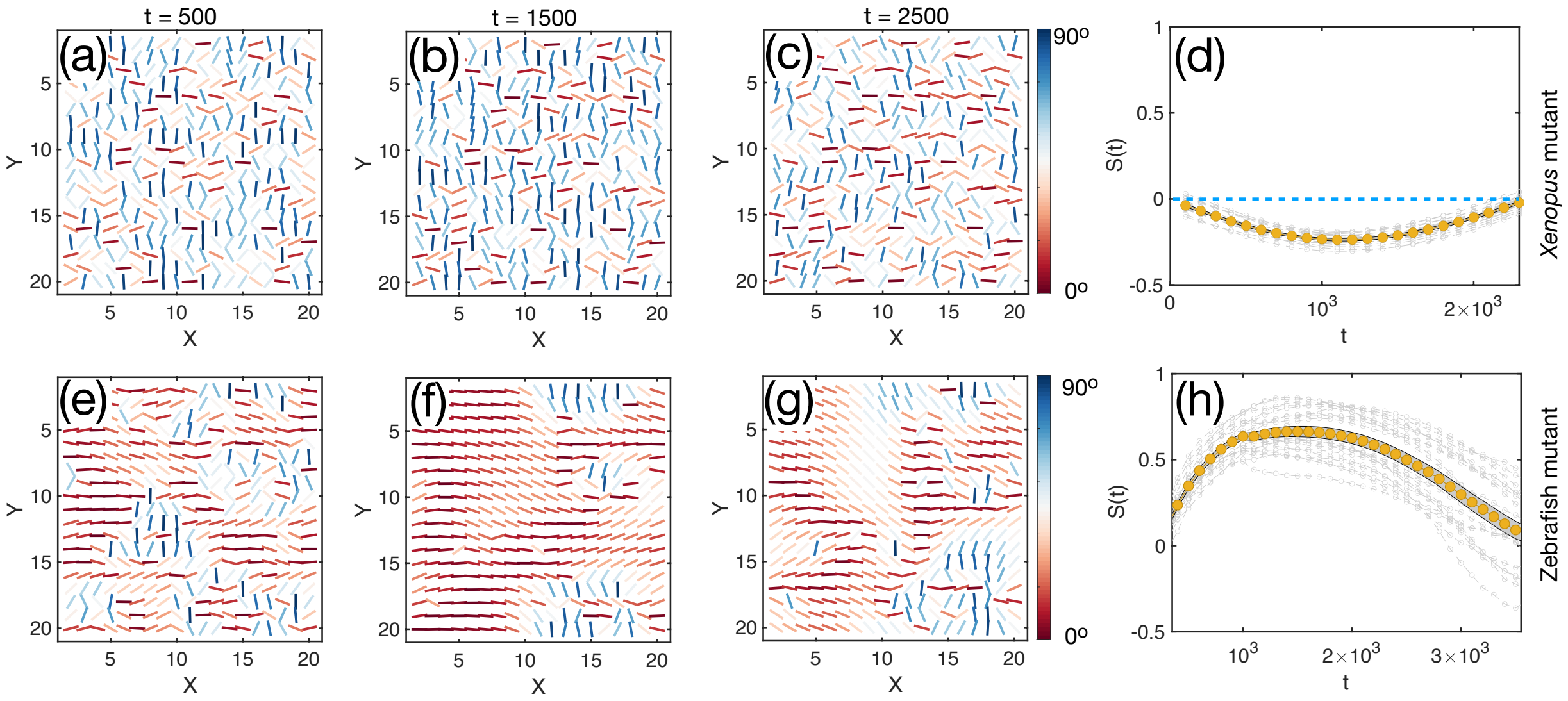

Figure 9. **Computational results for the mutant *Xenopus* and zebrafish tissues.** (a)-(c) The cell orientation (θ) in *Xenopus* mutants at different times. The short lines are color coded by the value of |θ|. The parameter values are: $\mathcal{A} = 10^{-6}$, and $\mathcal{B} = 3\times10^{-4}$ (see Eq. (7) in the main text). Time is measured in simulation steps. (d) The temporal evolution of the nematic order parameter S(t) of cells in mutant *Xenopus* tissues. Each curve in gray represents a single trajectory. Data in the brown dot is the mean value averaged over 20 trajectories. The shaded area shows the SEM. (e)-(h) Similar to (a)-(d) for zebrafish mutant tissues. The parameter values are: $\mathcal{A} = 2.5\times10^{-3}$, and $\mathcal{B} = 1\times10^{-3}$ for $t < 1000$, while $\mathcal{B} = -1\times10^{-3}$ for $t \geq 1000$.

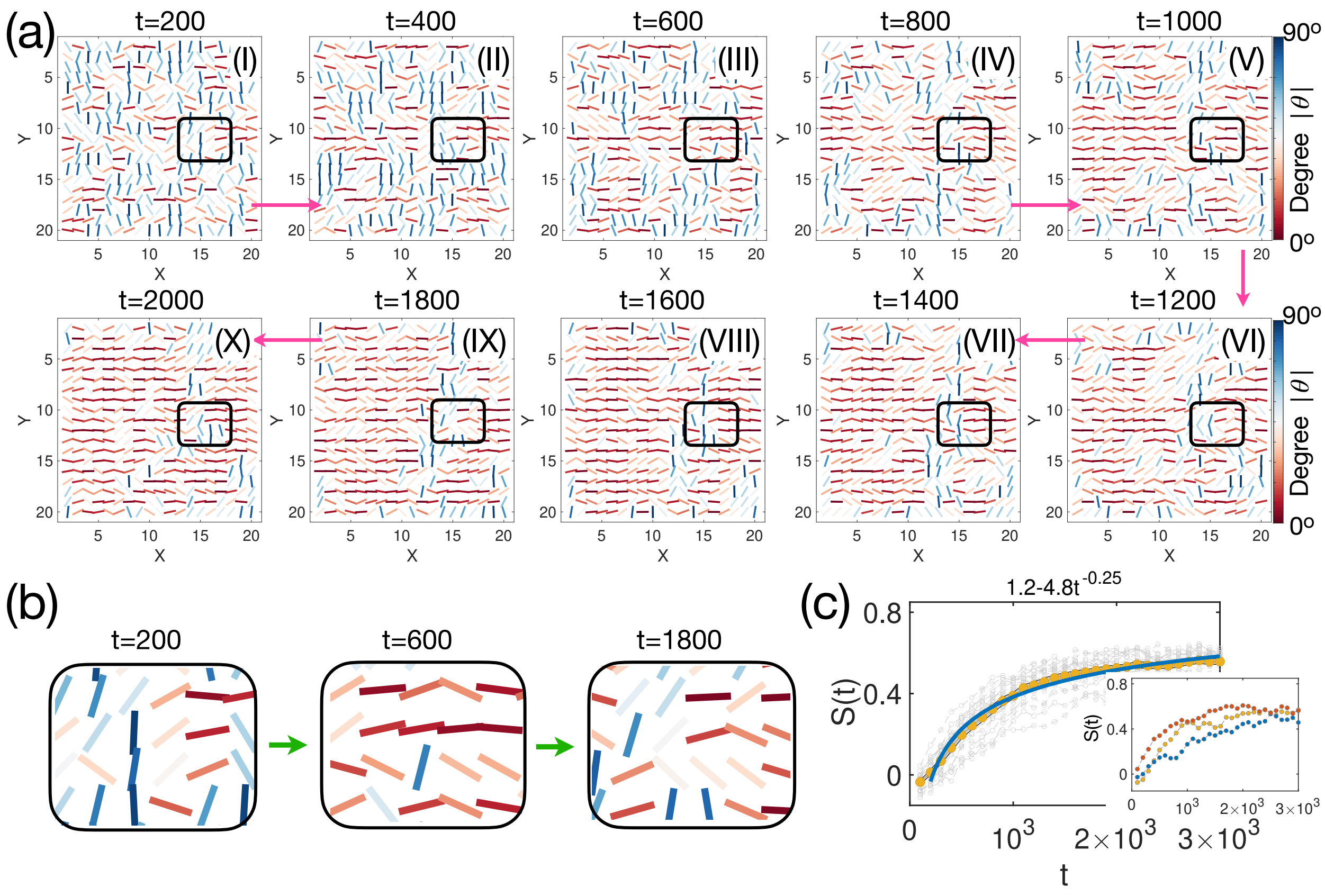

*Figure 10.* **Nematic order in the presence of noise.** (a) Cell orientation at different times t. The parameters $\mathcal{A}$ and $\mathcal{B}$ are the same as used in model (iii) in Fig. 4(i)-(l) but with an additional term representing the noise, $C\zeta(t)$. $\zeta$ is white noise with zero mean and variance, $\langle\zeta(t)\zeta(t')\rangle = \delta(t - t')$. The value of C is $5 \times 10^{-2}$. Like Fig. 4(i)-(k), cells become orientated along the horizontal axis as time t increases. Certain regions (see the one enclosed by the black boxes in each figure for an example) can change from disorder to order or vice versa due to noise-induced fluctuations. (b) A zoomed in view of the region in the boxes for (a) to showing the transitions between the ordered and disordered phases. (c) Time dependent changes in $S$ for all cells in (a). Each curve in gray represents one realization. Data in the brown dot is the mean value averaged over 20 trajectories. The shaded area shows the SEM. Navy blue line shows a power-law behavior with the function listed at the top of the figure. The inset shows several representative individual trajectories from the main figure, showing a fluctuation of $S$ over time.

# Emergence of cellular nematic order is a conserved feature of gastrulation in animal embryos - Supplementary Information

Xin Li,[1, a)] Robert J Huebner,[2, b)] Margot L.K. Williams,[3,4] Jessica Sawyer,[5,6] Mark Peifer,[6] John B Wallingford,[2] and D. Thirumalai[1,7]

[1)] *Department of Chemistry, University of Texas at Austin, Austin, TX 78712, USA*

[2)] *Department of Molecular Biosciences, University of Texas at Austin, Austin, TX 78712, USA*

[3)] *Center for Precision Environmental Health & Department of Molecular and Cellular Biology, Baylor College of Medicine, Houston, TX 77030, USA*

[4)] *Department of Developmental Biology, Washington University School of Medicine, St. Louis, MO 63110, USA*

[5)] *Department of Pharmacology and Cancer Biology, Duke University, Durham, NC 27710, USA*

[6)] *Department of Biology, University of North Carolina at Chapel Hill, Chapel Hill, NC 27599-3280, USA*

[7)] *Department of Physics, University of Texas at Austin, Austin, TX 78712, USA*

[a)]Electronic mail: xinlee0@gmail.com

[b)]Electronic mail: roberth@utexas.edu

In the Supplementary Information (SI), we provide additional figures and explanations for the Videos that are pertinent to the results present in the main text.

**SI Figures**

Figure S1. **Snapshots of the mesoderm layer of zebrafish tissues and spatial-temporal evolution of the cell shape index** $SI$ **during** $CE$**.**

Figure S2. **Evolution of the cell area distribution in zebrafish.**

Figure S3. **Shape index along the anteroposterior axis at different times.**

Figure S4. **Evolution of the cell shape index outside the notochord.**

Figure S5. **Nematic order during zebrafish development.**

Figure S6. **Spatial-temporal evolution of nematic order during zebrafish development.**

Figure S7. **Evolution of cell orientation during zebrafish** $CE$**.**

Figure S8. **T1 transitions in zebrafish** $CE$**.**

Figure S9. **Nucleation and expansion of the nematic phase during zebrafish** $CE$**.**

Figure S10. **Cell morphology changes during** *Xenopus* $CE$**.**

Figure S11. **Calculated spatial correlation function** $C_S(r)$ **(see Eq. 2 in the main text) using Model III**.

Figure S12. **Comparison of cell shape index for wild-type** *Xenopus* **versus** *C-cadherin* **knockdown tissues.**

Figure S13. **Calculated nematic order formation using simulations with lattice size 30×30.**

Figure S14. **Computational results using lattice size 50×50.**

**SI Table**

Table I. **The parameters used in the simulation.**

**SI Videos**

Video I. **Movie for zebrafish** $CE$**.**

Video II. **The evolution of the cell shape index (**$SI$**) as a function of the cell position (**$x$**) along the mediolateral axis of zebrafish in Video 1.**

Video III. **The evolution of the cell orientation in zebrafish *CE*.**

Video IV. **Movie for *Xenopus laevis CE*.**

Video V. **Movie for *Drosophila CE*.**

Video VI. **Movie for *Xenopus laevis CE* after PCP-protein knockout.**

Video VII. **Movie for *Xenopus laevis CE* after Cdh3 knockout.**

Video VIII. **Movie for zebrafish *CE* with spt/Tbx16 mutant.**

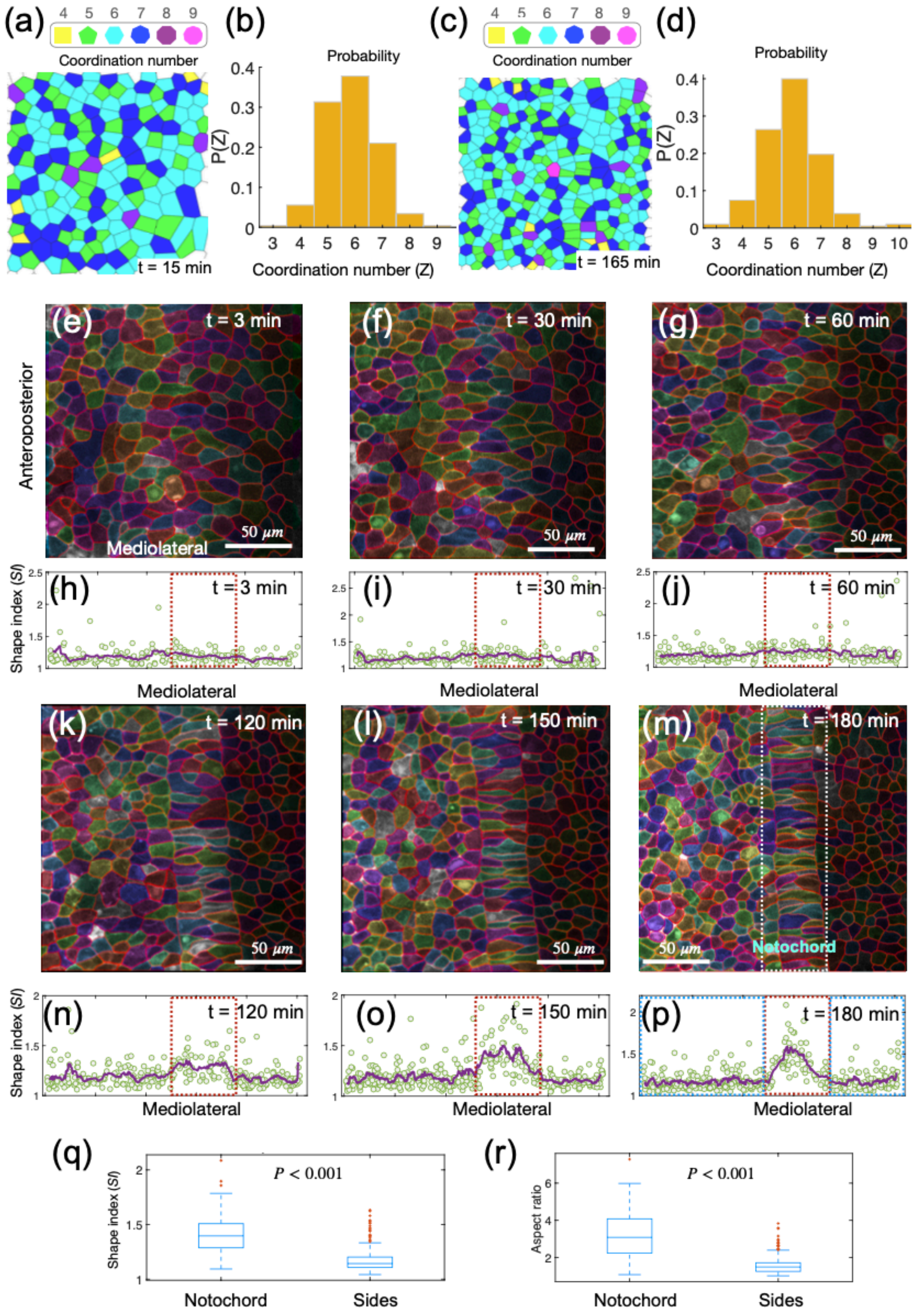
(a)
4 5 6 7 8 9
Coordination number
t = 15 min
(b)
Probability
P(Z)
Coordination number (Z)
(c)
4 5 6 7 8 9
Coordination number
t = 165 min
(d)
Probability
P(Z)
Coordination number (Z)
(e) t = 3 min
Anteroposterior
Mediolateral
50 μm
(f) t = 30 min
(g) t = 60 min
(h) t = 3 min
(i) t = 30 min
(j) t = 60 min
Shape index (SI)
Mediolateral
(k) t = 120 min
(l) t = 150 min
(m) t = 180 min
Notochord
(n) t = 120 min
(o) t = 150 min
(p) t = 180 min
(q)
P < 0.001
Shape index (SI)
Notochord
Sides
(r)
P < 0.001
Aspect ratio
Notochord
Sides

FIG. S1: **Snapshots of the mesoderm layer of zebrafish tissues and spatiotemporal evolution of the cell shape index $SI$ during $CE$.** (a), (c) Snapshots of the mesoderm layer of zebrafish at different timepoints, with cells segmented, and cell coordination number Z (the number of neighbors), derived from Voronoi tessellation of the cell centers. Each Voronoi cell is displayed in a different color based on the value of Z, as listed by the polygon symbols at the top of (a), (c). (b), (d) The distribution of Z for cells in (a), and (c). (e)-(g) and (k)-(m) Images of the zebrafish mesoderm layer at different times. Cell colors are for illustration purposes. (h)-(j) and (n)-(p) Shape index, $SI \equiv \frac{P}{\sqrt{4\pi A}}$, as a function of the cell position along the mediolateral axis corresponding to (e)-(g) and (k)-(m). The green dots represent different cells along the anteroposterior axis with a certain x-value (mediolateral axis). The purple line is the average of these points for smoothing the data. (q) Box-plot of the $SI$ of cells located in the notochord region (the red dashed box) versus cells at the two sides (the blue dashed box) in (p). (r) Same as (q), except showing the cell aspect ratio. The two-sided Mann-Whitney U test is used for the statistical analysis. The p-value is listed in (q) and (r).

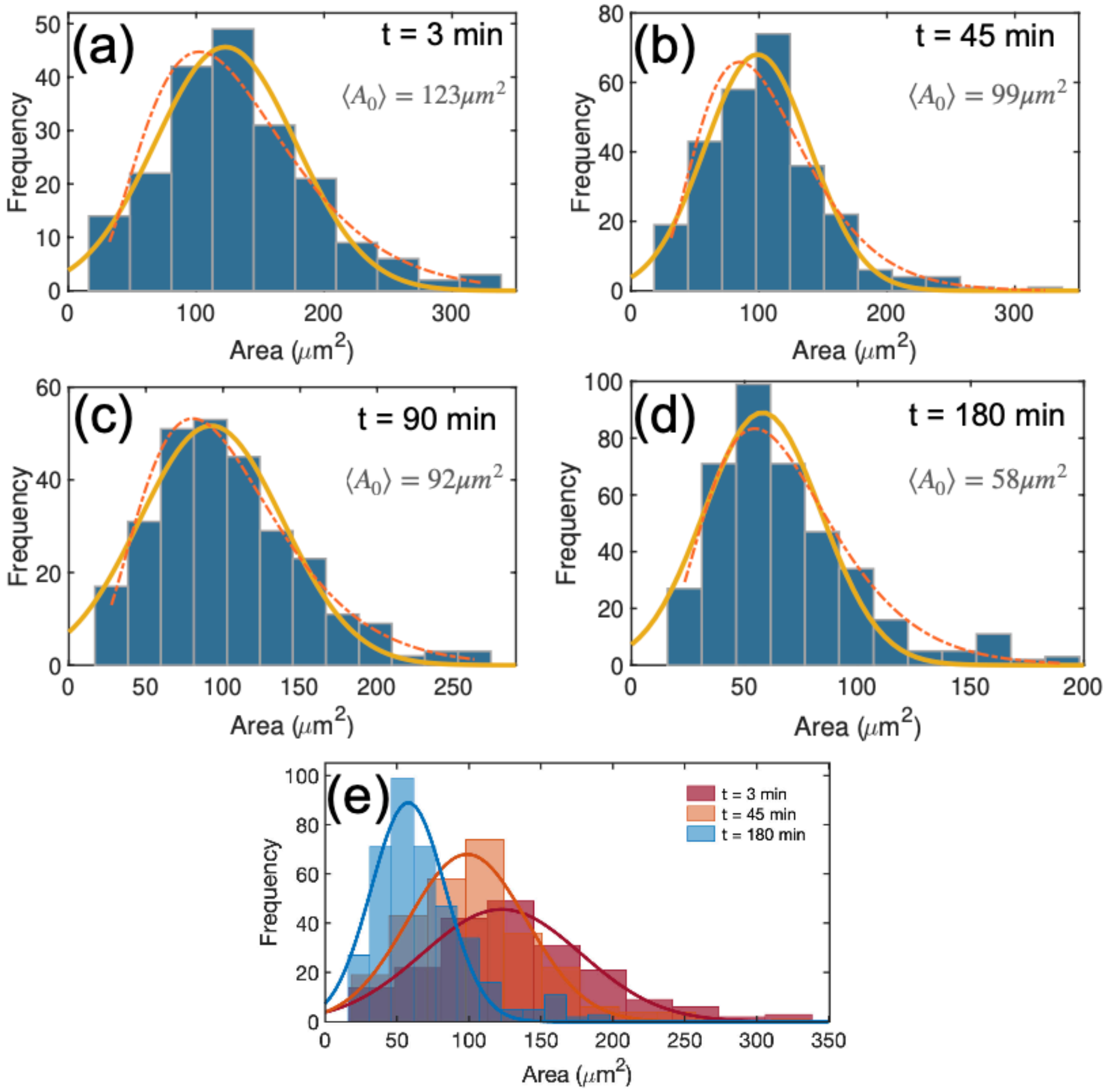

FIG. S2: **Evolution of the cell area distribution in zebrafish.** (a)-(d) Time dependent changes in the cell area distribution. (a)-(d) Histogram of the apical cell area at different times during zebrafish CE. The solid (dash-dotted) line is the Gaussian (Gamma) distribution fit to the data. The mean area, $\langle A_0 \rangle$, of cells from the Gaussian fit is listed in each figure. (e) Distribution of cell area at three time points. There is a shift towards the small values of the mean area as time increases.

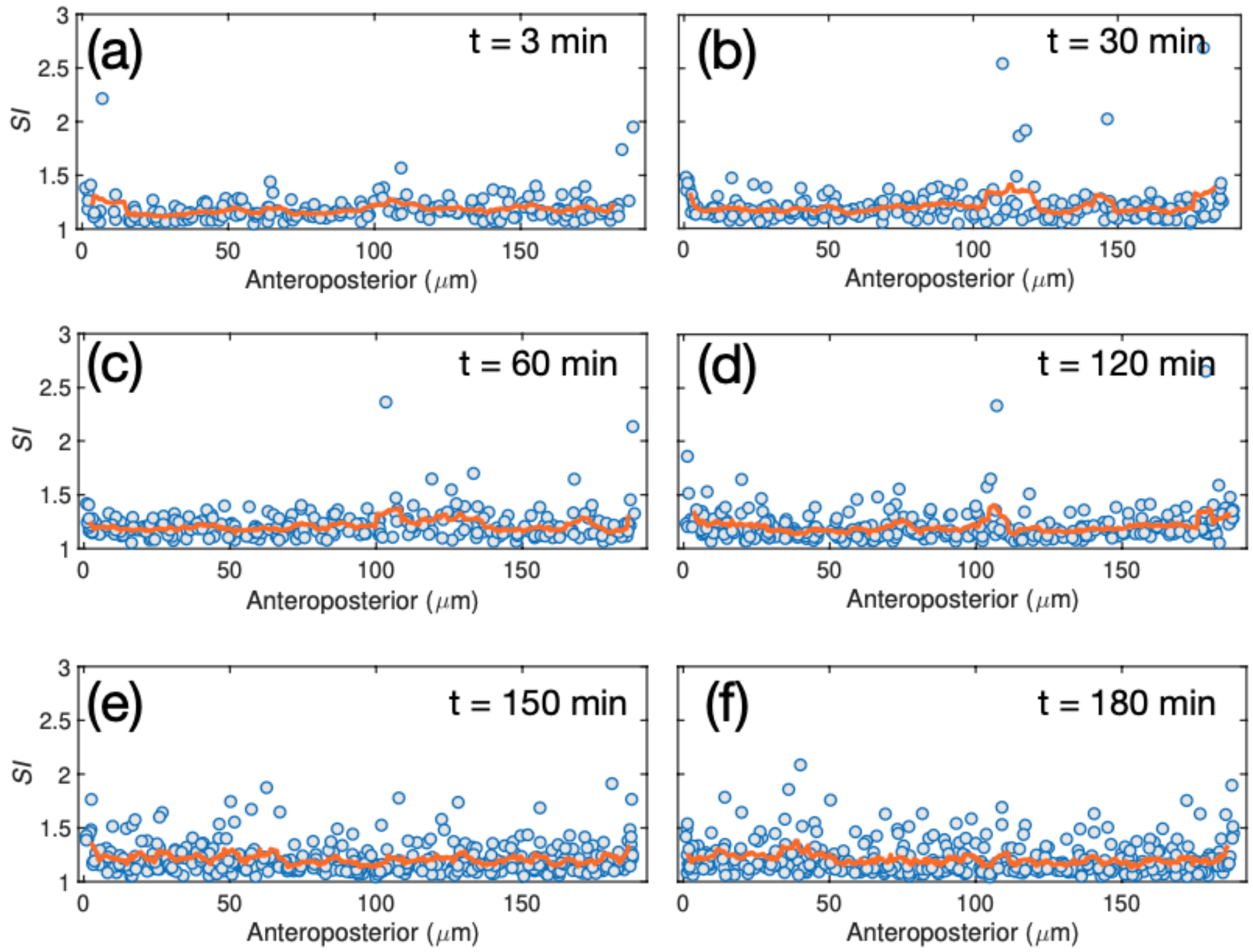

FIG. S3: **Shape index along the anteroposterior axis at different times.** (a)-(f) Same as Figs. S1(h)-(j) and (n)-(p), except for the value of $SI$ as a function of the cell position along the anteroposterior axis corresponding to Figs. S1(e)-(g) and (k)-(m).

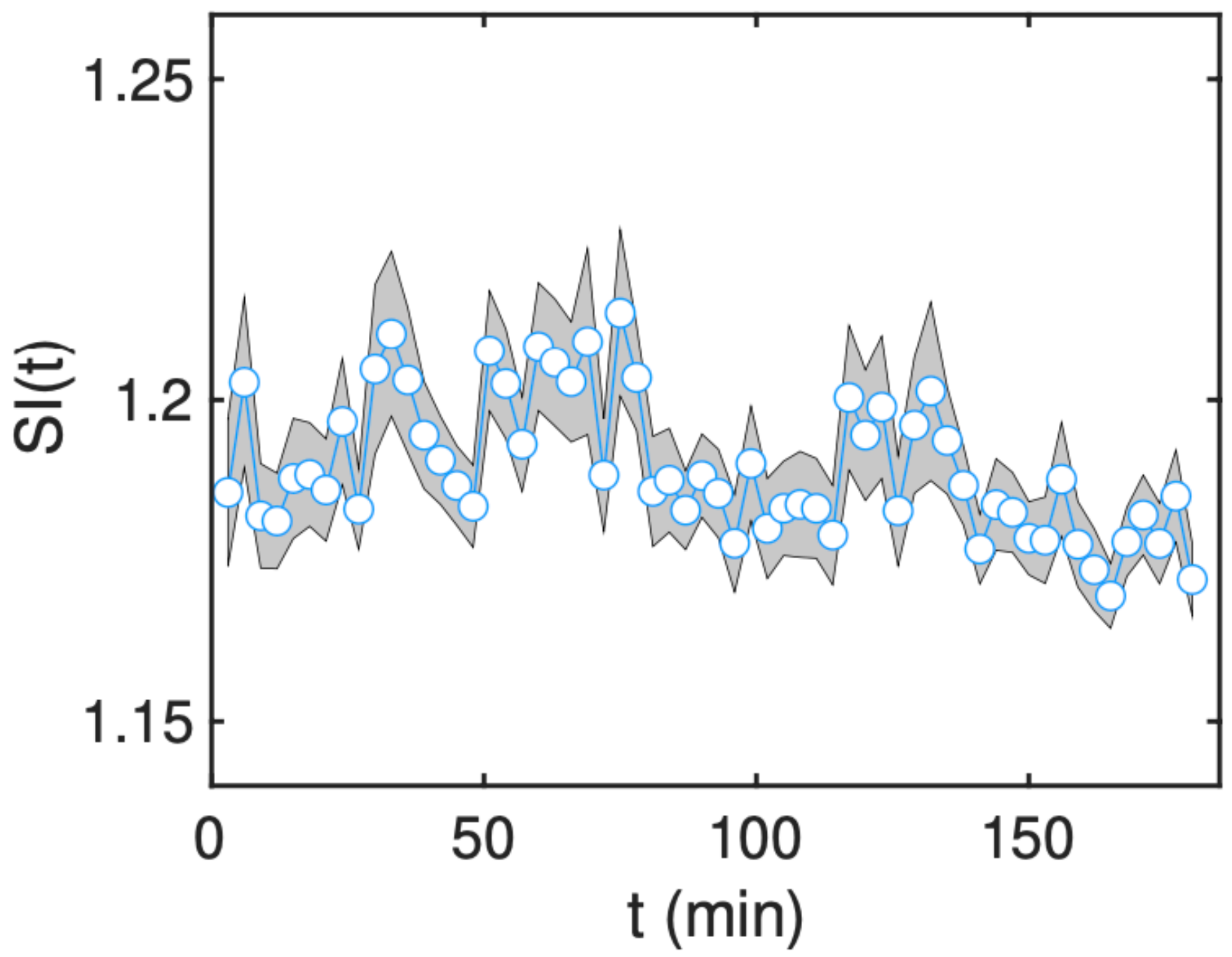

FIG. S4: **Evolution of the cell shape index outside the notochord.** Temporal evolution of the shape index, SI(t), of cells outside the notochord region. No significant change is observed in the time dependence of SI in this region.

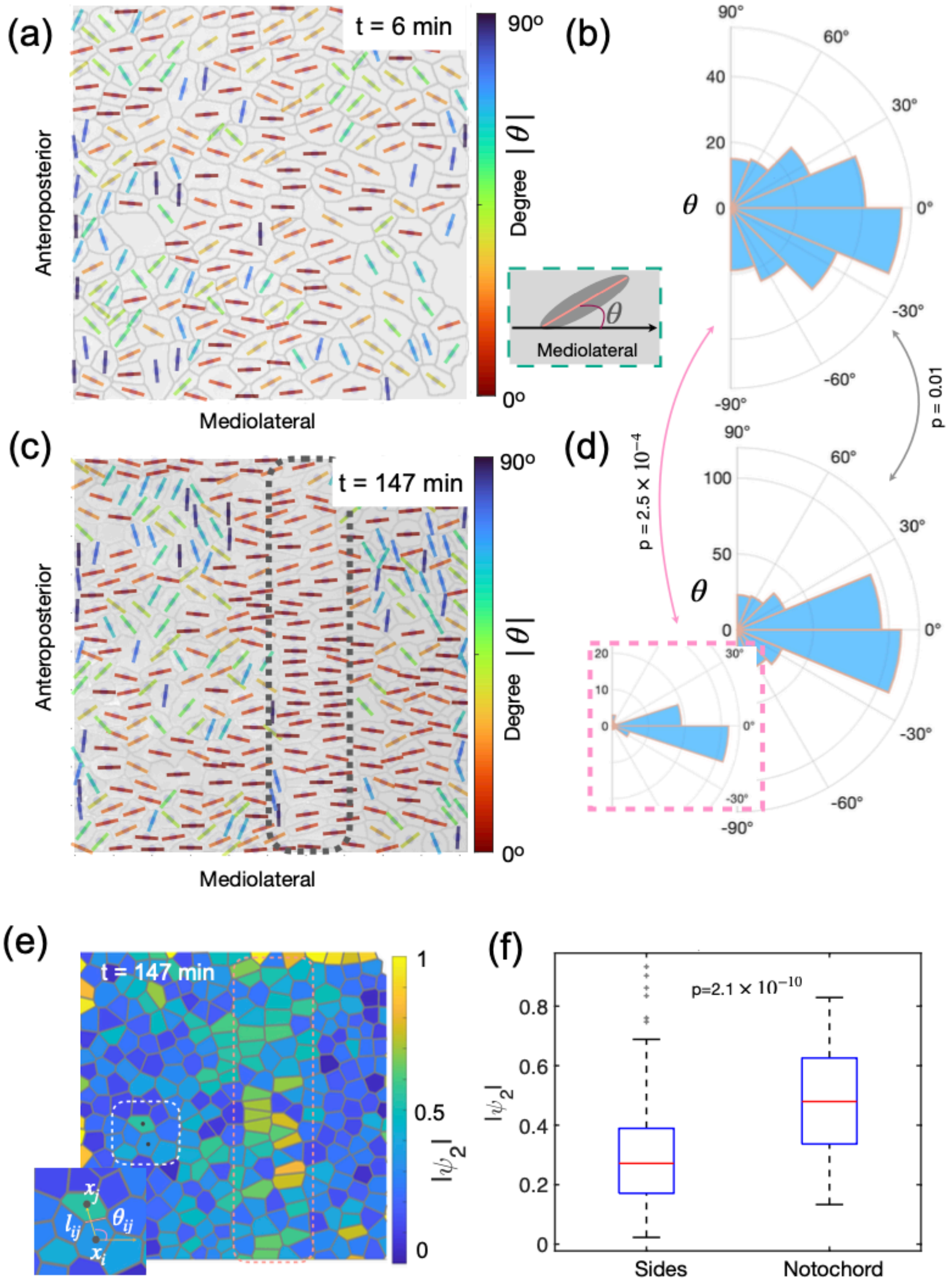

FIG. S5: **Nematic order during zebrafish development.**

FIG. S5: **Nematic order during zebrafish development.** (a) and (c) Same as Figs. 1(d) and (g) in the main text, showing the cell orientation in zebrafish tissues at different times. The cell orientation is defined by the angle, $\theta$, between the long axis of cells (see the short lines) and the horizontal (mediolateral) axis, as shown in the inset at the bottom right of (a). The short lines are color coded by the value of $|\theta|$. (b) Histogram of the polar angle $\theta$ for all cells in (a). (c)-(d) Same as (a)-(b), except they are at a later timepoint. The inset in (d) gives the histogram of $\theta$ for cells located in the notochord region shown by the pink dotted rectangle in (c). The p-values between the patterns in (b) and (d), (b) and the inset in (d) are calculated with two-sample Kolmogorov–Smirnov test. (e) The 2-fold orientation order, $|\psi_2|$, of cells (see the definition in Eq.(3) in the Materials and Methods section). Each Voronoi cell is color coded by the value of $|\psi_2|$ (see the scale on the right). The inset on the left bottom is a zoom in of the white dashed rectangle region. The variables $\theta_{ij}$ (the angle between the vector pointing from cell i to j and the horizontal axis) and $l_{ij}$ (the length of the edge shared between the Voronoi cells i and j) are used in the definition of $|\psi_2|$. (f) Boxplot of $|\psi_2|$ for cells located at two sides or the middle region (see the dotted rectangle in (e)). The p-value is calculated using the two-sided Mann-Whitney U test.

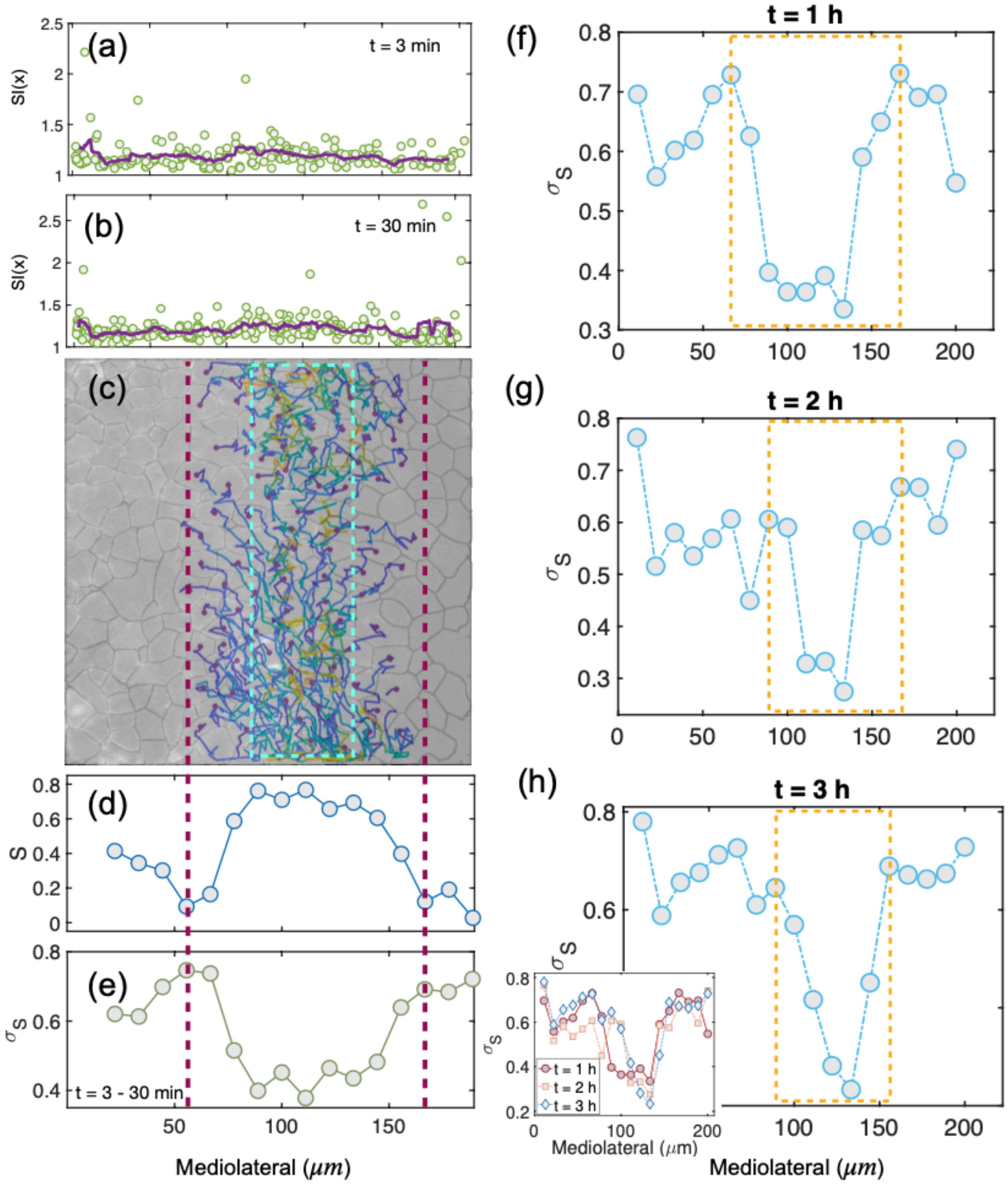

FIG. S6: **Spatiotemporal evolution of nematic order during zebrafish development.**

FIG. S6: **Spatio-temporal evolution of nematic order during zebrafish development.** (a)-(b) The cell shape index, $SI(x)$, along the mediolateral axis at early times. (c) The tissue snapshot at an early time point is overlaid with the trajectories of cells located inside of the red dashed lines over 3 hours, with blue color for early time points and yellow for late timepoints. The dashed rectangle in cyan gives the final notochord region. (d)-(e) The nematic order parameter, S, and the standard deviation, $\sigma_S$, along the mediolateral axis. The value of S and $\sigma_S$ in (d)-(e) are averaged over 10 successive time frames (t=30 minutes). Note the fluctuations in the notochord is significantly less than for cells that are outside it. The boundaries between the nematic-order and disordered regions are indicated by the red dashed lines. (f)-(h) The spatial-temporal evolution of $\sigma_S$. A sharp transition of $\sigma_S$ along the mediolateral axis is observed as indicated by the dashed rectangles at different times, and this region narrows over time. The inset in (h) shows the value of $\sigma_S$ at different times.

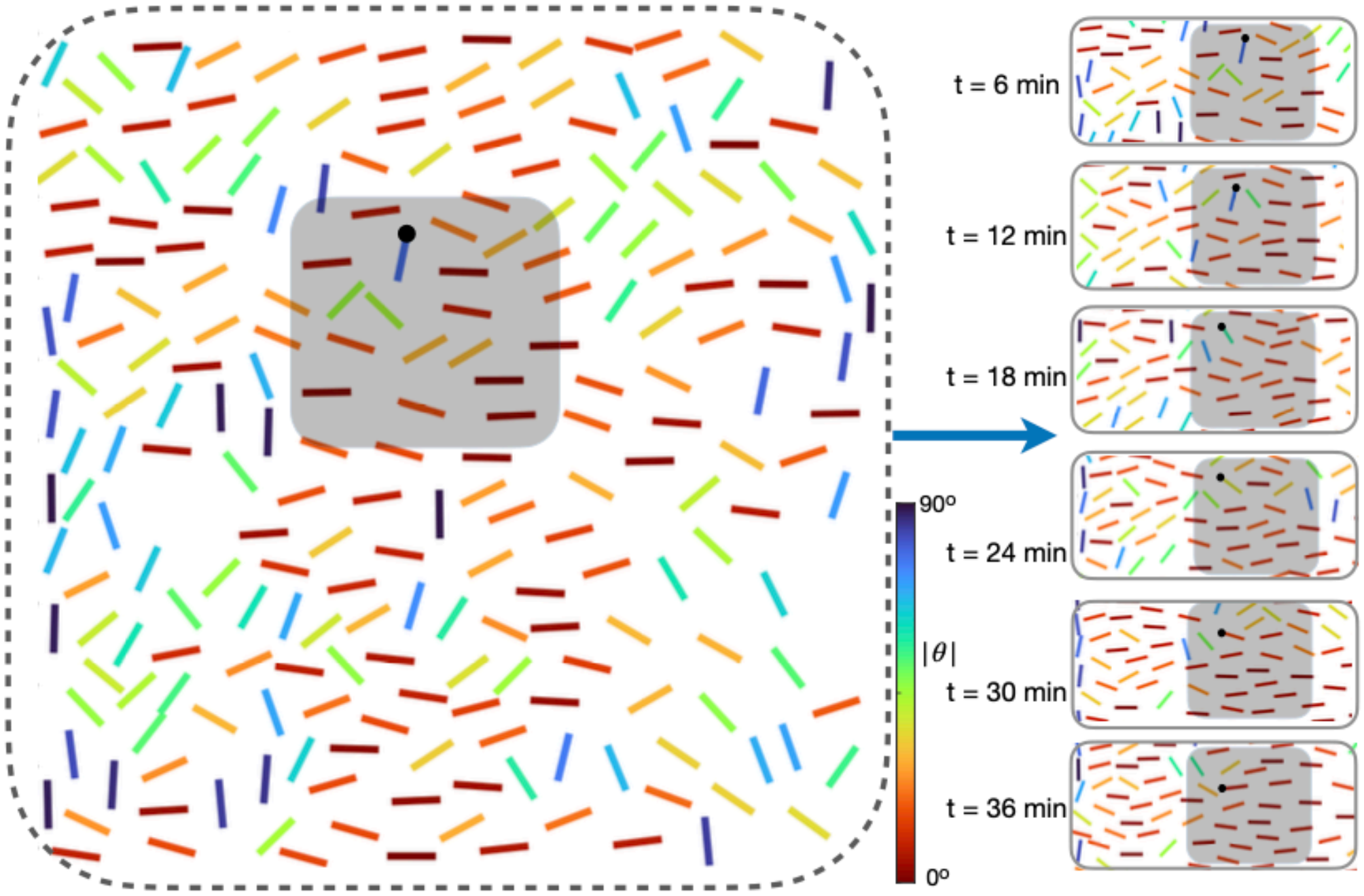

FIG. S7: **Evolution of cell orientation during zebrafish** ***CE*****.** Similar to Fig. 1(d) in the main text. The cell orientation is defined by the angle, $\theta$, between the long axis of cells (see the short lines) and the mediolateral (horizontal) axis as shown in the inset in Fig. S5(a). The lines are color coded by the value of $|\theta|$. Figures on the right show the evolution of cell orientation located in the shaded area of the left figure. There are dramatic changes in the orientation of individual (see the line with a dot on one side as an example) in the process of forming a nematic phase.

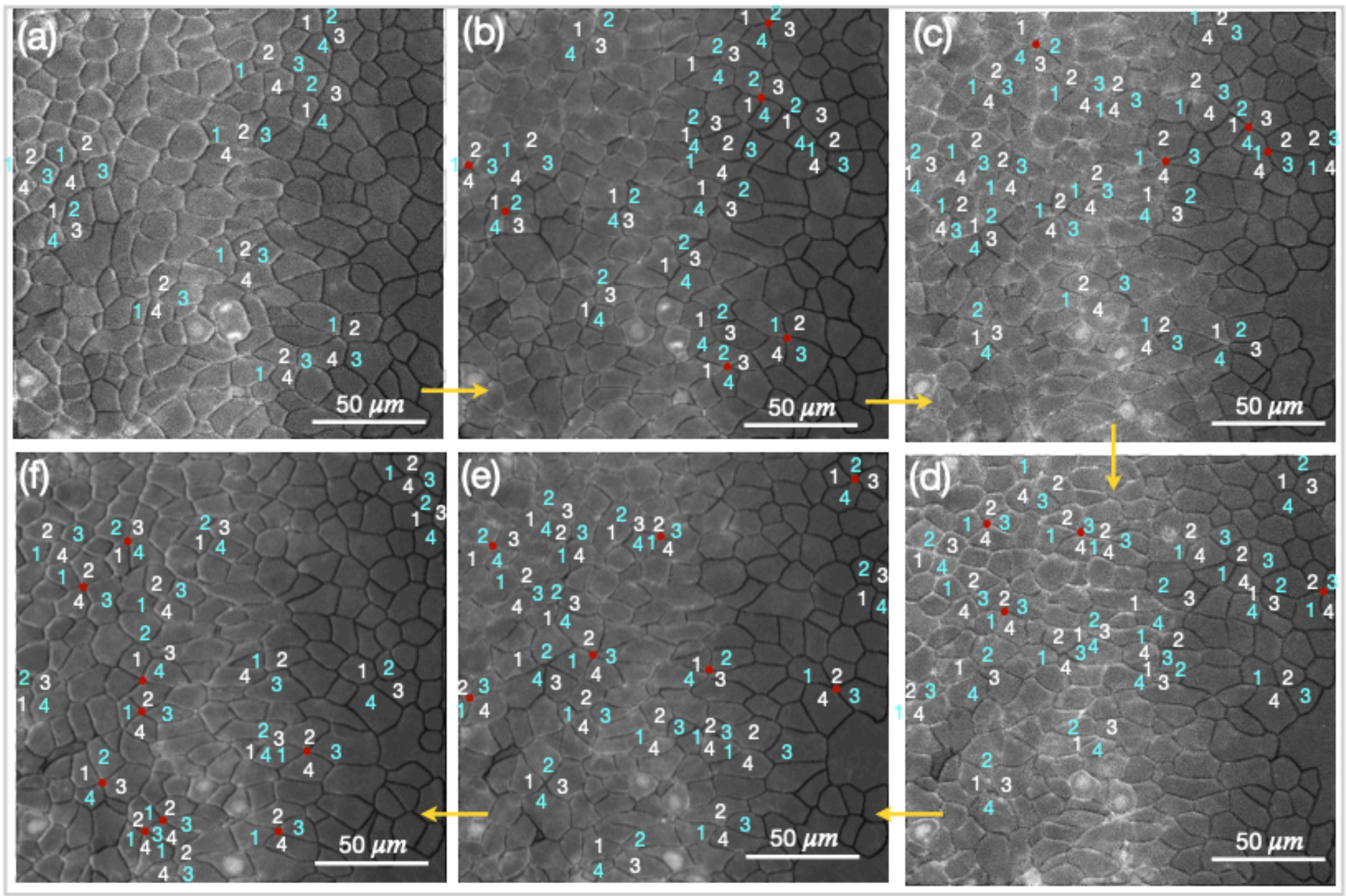

FIG. S8: **T1 transitions in zebrafish** ***CE*****.** Several examples of T1 transitions in zebrafish *CE* in six consecutive images. Four cells are labeled 1, 2, 3 and 4 with white color indicating two cells that are in contact with each other while cyan color indicates they are separated. The red dot shows where two vertices meet.

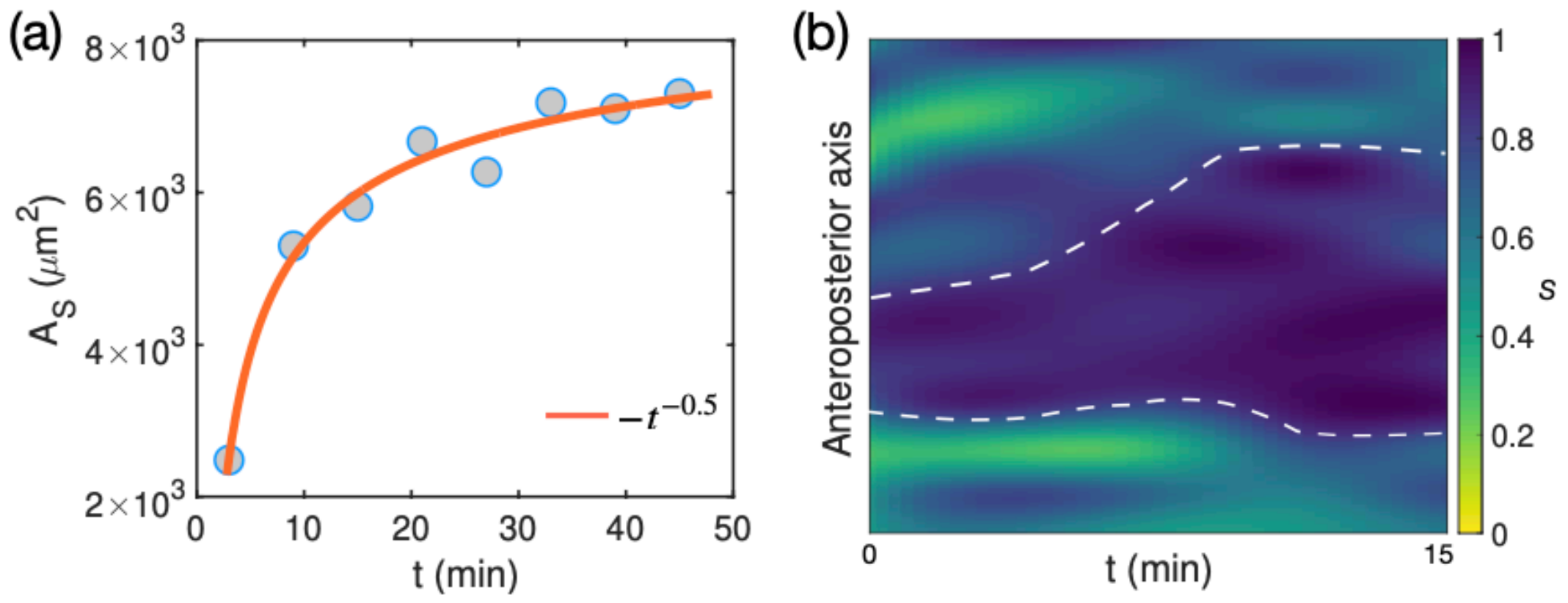

FIG. S9: **Nucleation and expansion of the nematic phase during zebrafish** ***CE*****.** (a) The area of the nematic phase, $A_S$ (with $S > 0.8$), as a function of time in the notochord region corresponding the Fig. 1l in the main text. (b) A kymograph showing the mean values of S (averaged over the mediolateral axis on the notochord region) along the anteroposterior axis at different times. The dashed lines (indicating the boundaries of the nematic order region) are a guide to the eye, showing the nucleation and growth of the nematic order.

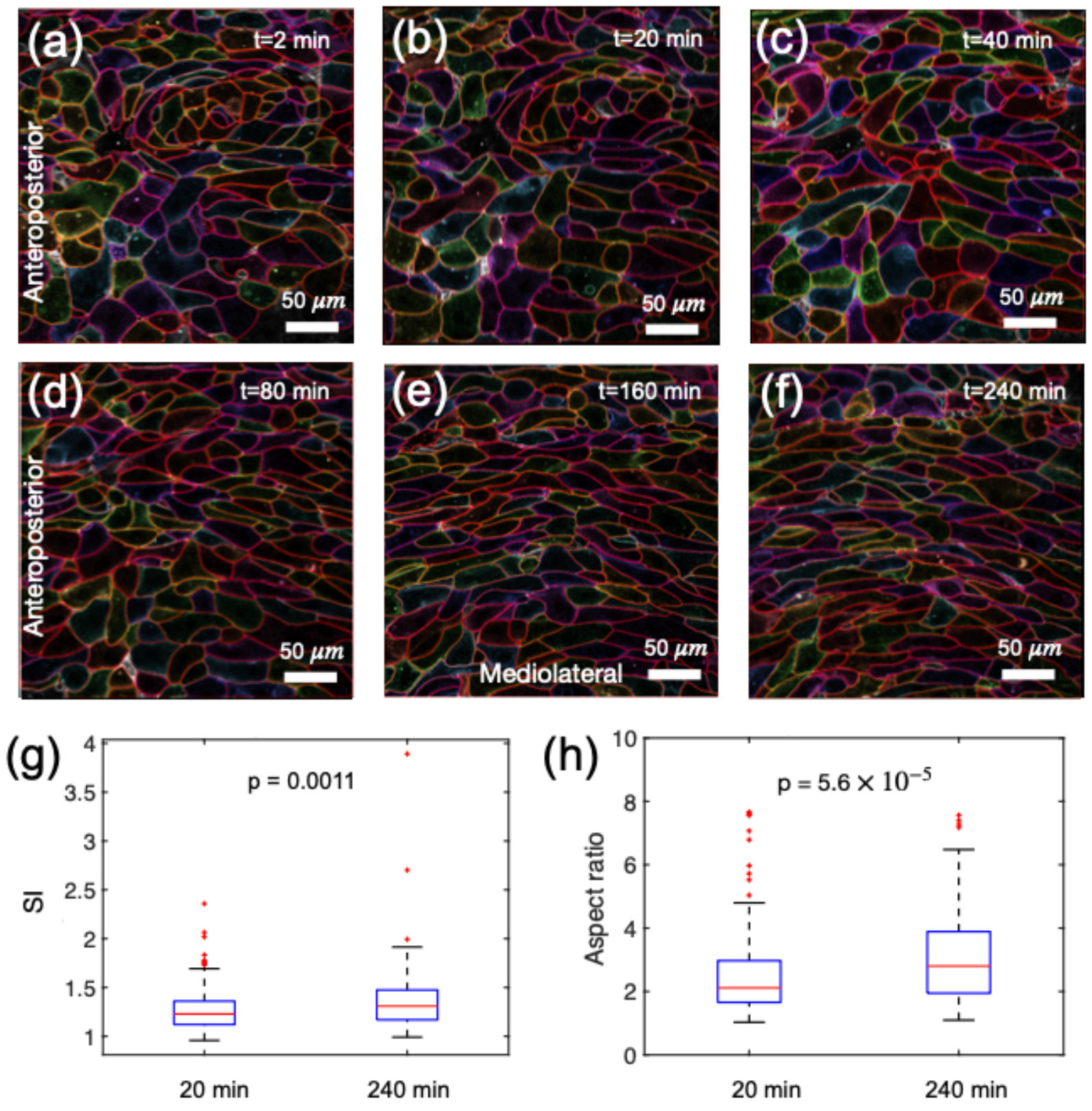

FIG. S10: **Cell morphology changes during** ***Xenopus CE*****.** (a)-(f) Snapshots of *Xenopus laevis* tissues at different times during *CE*. (g) Box-plot of the *SI* value calculated for cells from (b) and (f). (h) Aspect ratio of cells from (b) and (f). The two-sided Mann-Whitney U test is used for the statistical analysis in (g)-(h). The p-value is listed in each figure.

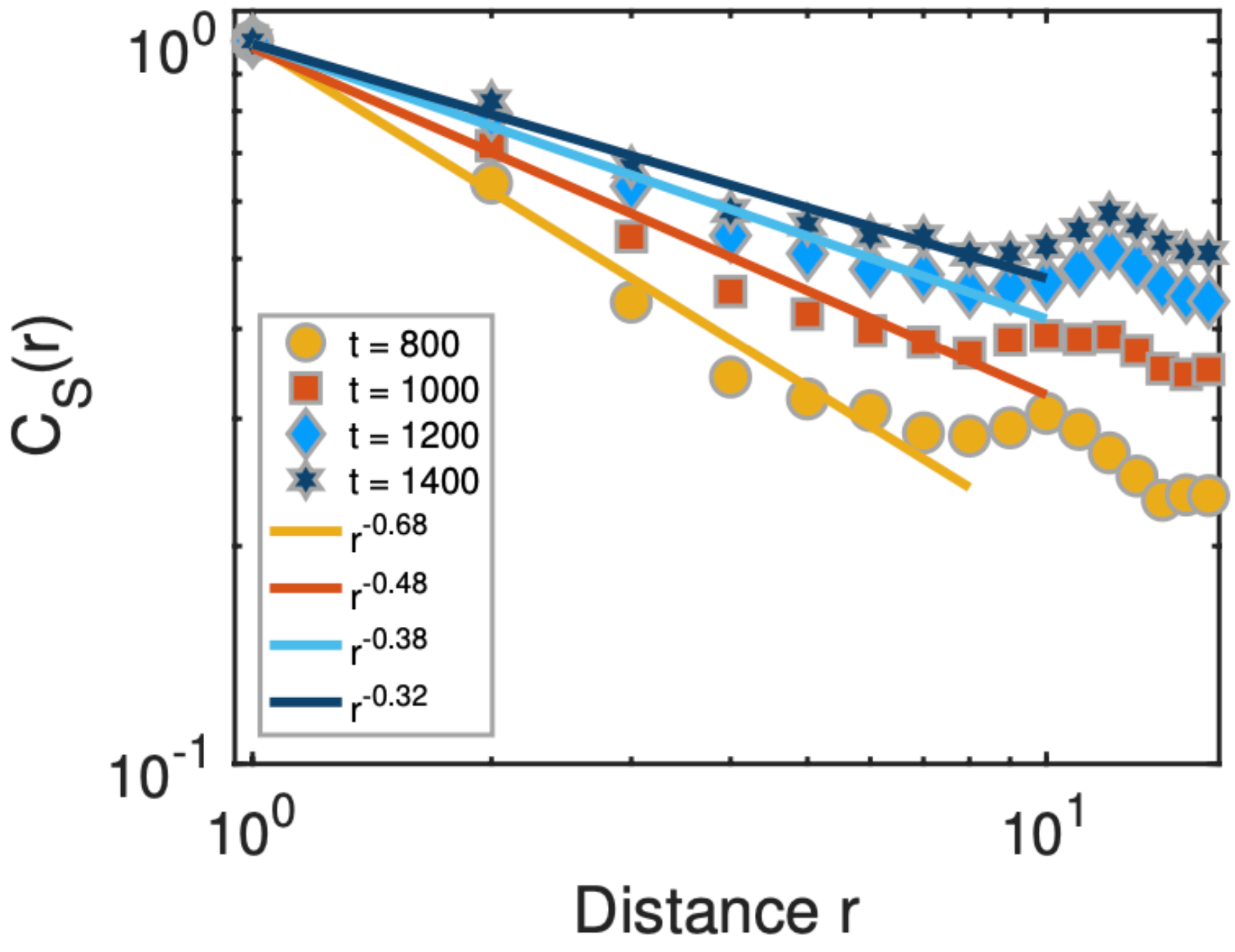

FIG. S11: **Calculated spatial correlation function $C_S(r)$ (see Eq. 2 in the main text) using Model III**. The spatial correlation, $C_S(r)$, of cells at different times from simulations in Model III with the same parameters used in Fig. 4i-l in the main text. The solid lines are the power-law fits to the data.

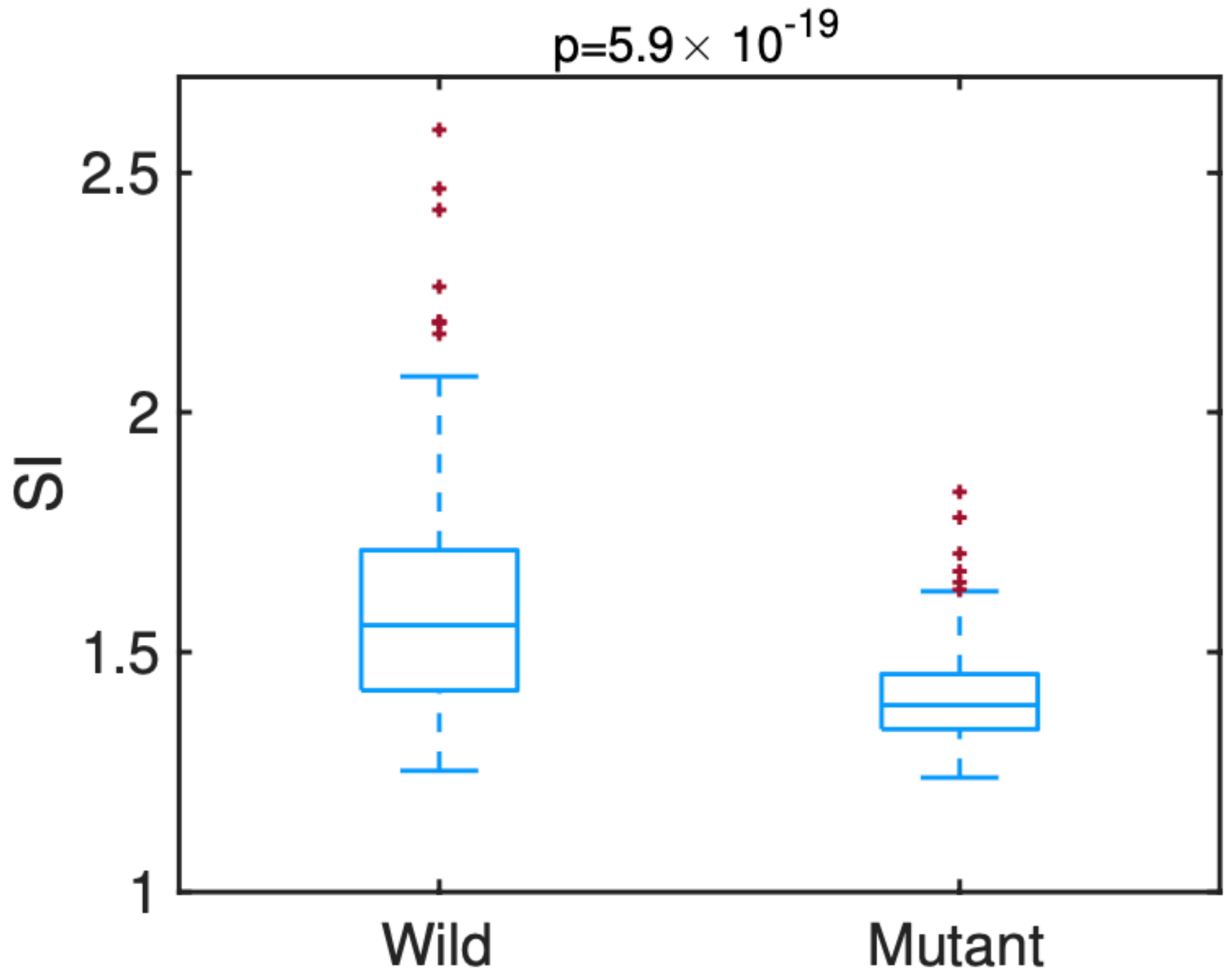

FIG. S12: **Comparison of cell shape index for wild-type** *Xenopus* **versus** *C-cadherin* **knockdown tissues.** Box-plot of the $SI$ for cells from Figs. 6(a) and (b) in the main text for *C-cadherin* knockdown and wild-type *Xenopus* tissues. Mann-Whitney U test is used for the statistical analysis. The p-value is listed in the figure.

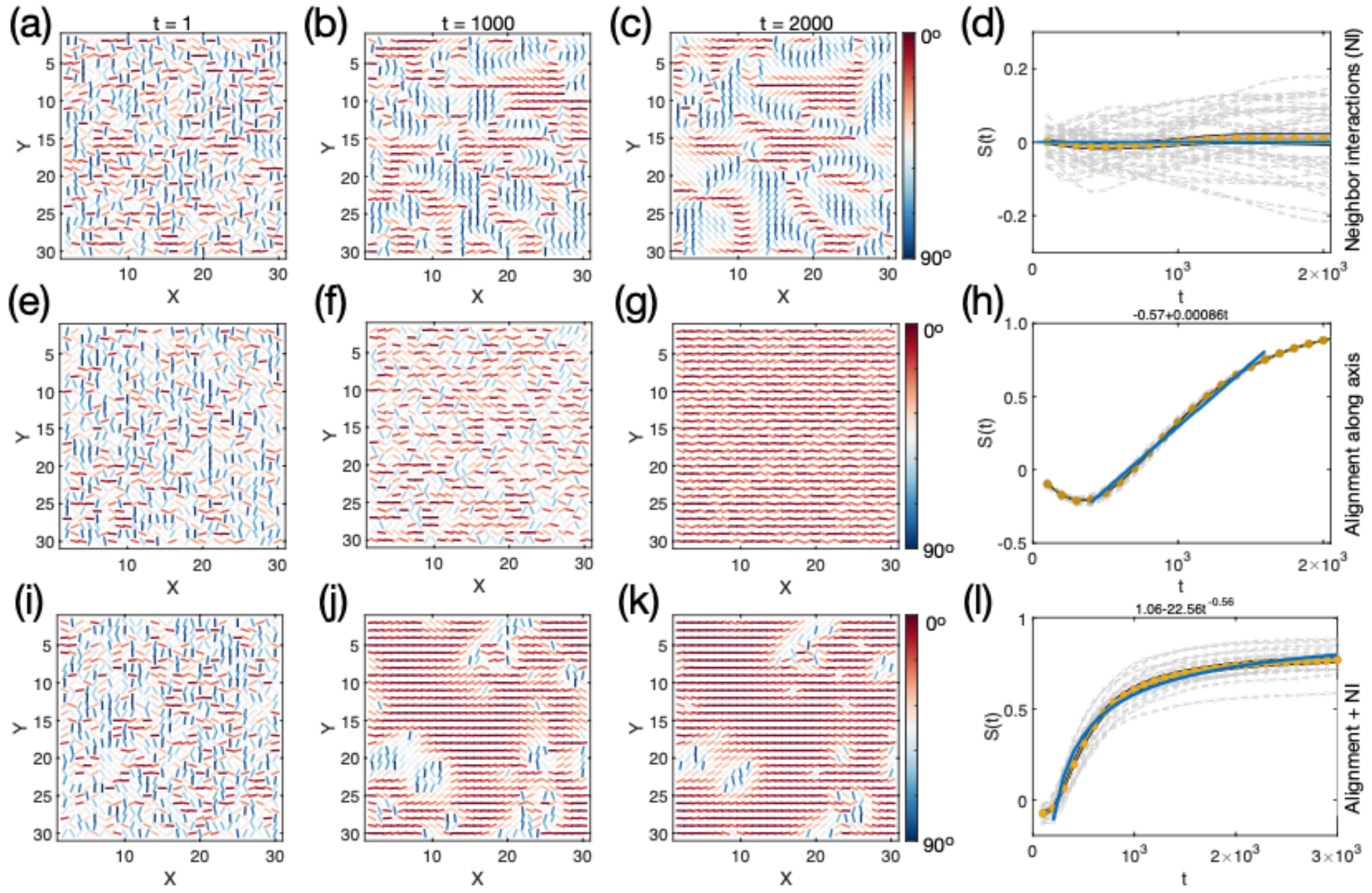

FIG. S13: **Calculated nematic order formation using simulations with lattice size 30×30.** Same as Figure 4 in the main text, except the lattice size is 30×30. (a)-(d) Results from model (i). (e)-(h) Results from model (ii). (i)-(l) Results from model (iii). The parameter $\mathcal{A}$=2.5×10$^{-3}$, $\mathcal{B}$=1×10$^{-3}$. The models are described in the main text.

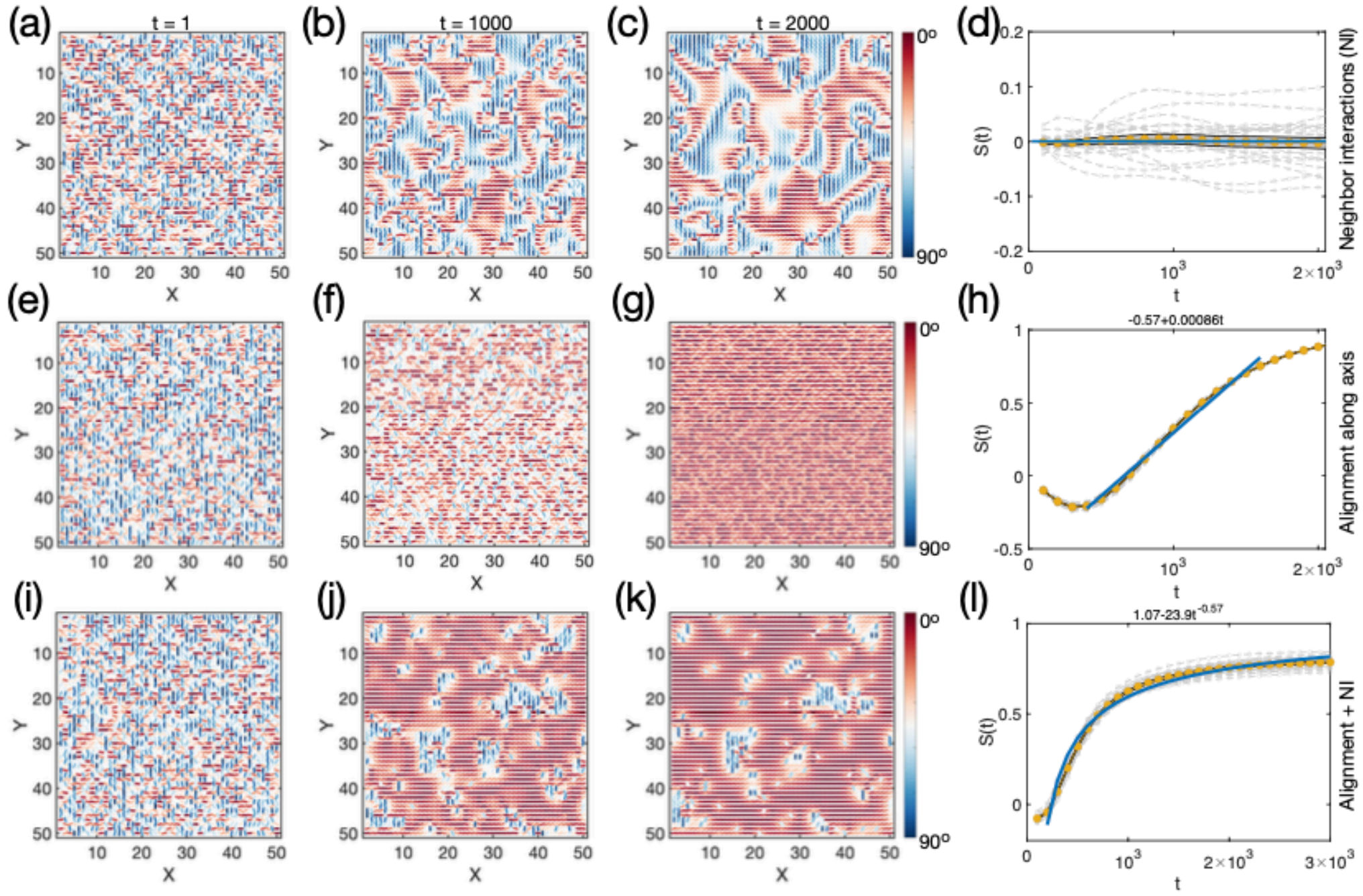

FIG. S14: **Computational results using lattice size 50×50.** Same as Figure S13, except the lattice size is 50×50.

TABLE I: The parameters used in the simulation.

| **Models** | $\mathcal{A}(local)$ | $\mathcal{B}(global)$ |
|---|---|---|
| Model i | $2.5\times10^{-3}$ | 0 |
| Model ii | 0 | $1\times10^{-3}$ |
| Model iii | $2.5\times10^{-3}$ | $1\times10^{-3}$ |
| Model iii (*Xenopus* mutant) | $10^{-6}$ | $3\times10^{-4}$ |
| Model iii (zebrafish mutant) | $2.5\times10^{-3}$ | $-1\times10^{-3}$ |